\documentclass[sigconf,table,xcdraw,authorversion]{acmart}

\AtBeginDocument{%
  \providecommand\BibTeX{{%
    \normalfont B\kern-0.5em{\scshape i\kern-0.25em b}\kern-0.8em\TeX}}}

\copyrightyear{2024}
\acmYear{2024}
\setcopyright{acmlicensed}\acmConference[CHI '24]{Proceedings of the CHI Conference on Human Factors in Computing Systems}{May 11--16, 2024}{Honolulu, HI, USA}
\acmBooktitle{Proceedings of the CHI Conference on Human Factors in Computing Systems (CHI '24), May 11--16, 2024, Honolulu, HI, USA}
\acmDOI{10.1145/3613904.3642934}
\acmISBN{979-8-4007-0330-0/24/05}
\newcolumntype{L}{>{\raggedright\arraybackslash}X}

\usepackage{subcaption}
\usepackage{framed}

\definecolor{myblue}{HTML}{94caec}
\definecolor{myred}{HTML}{c26a77}
\definecolor{mygreen}{HTML}{5da899}
\definecolor{myyellow}{HTML}{dcce7d}

\definecolor{myblue80}{HTML}{a9d5f0}
\definecolor{myblue55}{HTML}{c4e2f5}

\definecolor{myred80}{HTML}{ce8892}
\definecolor{myred55}{HTML}{ddadb4}

\begin{document}


\title{Unraveling the Dilemma of AI Errors}
\subtitle{Exploring the Effectiveness of Human and Machine Explanations for Large Language Models}

\author{Marvin Pafla}
\email{mpafla@uwaterloo.ca}
\affiliation{%
  \institution{University of Waterloo}
  \streetaddress{200 University Ave W}
  \city{Waterloo}
  \state{Ontario}
  \country{Canada}
  \postcode{N2L 3G1}
}

\author{Kate Larson}
\email{kate.larson@uwaterloo.ca}
\affiliation{%
  \institution{University of Waterloo}
  \streetaddress{200 University Ave W}
  \city{Waterloo}
  \state{Ontario}
  \country{Canada}
  \postcode{N2L 3G1}
}

\author{Mark Hancock}
\email{mark.hancock@uwaterloo.ca}
\affiliation{%
  \institution{University of Waterloo}
  \streetaddress{200 University Ave W}
  \city{Waterloo}
  \state{Ontario}
  \country{Canada}
  \postcode{N2L 3G1}
}








\renewcommand{\shortauthors}{Pafla, et al.}

\begin{abstract}
The field of eXplainable artificial intelligence (XAI) has produced a plethora of methods  (e.g., saliency-maps) to gain insight into artificial intelligence (AI) models, and has exploded with the rise of deep learning (DL). However, human-participant studies question the efficacy of these methods, particularly when the AI output is wrong. In this study, we collected and analyzed 156 human-generated text and saliency-based explanations collected in a question-answering task ($N=40$) and compared them empirically to state-of-the-art XAI explanations (integrated gradients, conservative LRP, and ChatGPT) in a human-participant study ($N=136$). Our findings show that participants found human saliency maps to be more helpful in explaining AI answers than machine saliency maps, but performance negatively correlated with trust in the AI model and explanations. This finding hints at the dilemma of AI errors in explanation, where helpful explanations can lead to lower task performance when they support wrong AI predictions. 



  
\end{abstract}



\maketitle

\section{Introduction}

With the rise of deep learning (DL) \cite{deep_learning} and large language models (LLMs), artificial intelligence (AI) has become ever more prominent in our lives, prompting a growing demand to make AI more explainable \cite{weld2018intelligibility, lipton_2018, miller2018explanation, kaur2020interpreting, yang2020reexamining}. To fill this demand, the field of XAI has produced numerous metrics and visualizations (e.g., \cite{samek2017explainable, ai_evaluation_feng2019, tutorials_lai}). However, there are doubts that these machine explanations, or XAI explanations (e.g., saliency maps), are actually able to explain LLMs (and other neural nets) and their behaviour \cite{miller2018explanation, weld2018intelligibility, lipton_2018}, and it is unclear how people interpret and use them. XAI has been criticized for cherry-picking explanations for correct predictions \cite{stop-Rudin-2019} and a lack of proper control conditions that compare XAI explanations with other or no explanations \cite{miller2018explanation, context-utility-vasilyeva-2017, goals-explanation-mohseni-2021}. There is a general lack of thorough human-participant studies \cite{miller2018explanation, meta-xai-schemmer-2022}.

To address existing limitations, this research empirically assesses state-of-the-art machine explanations for LLMs through a study involving human participants, with a specific focus on saliency maps as a primary mechanism for local explanations in LLMs. Despite previous studies on saliency maps and feature-based visualizations in classification tasks with incorrect predictions (e.g., \cite{Jacobs2021, trustcal2020zhang, whole2021bansal}), this investigation pioneers the examination of saliency maps in text generation tasks, considering both correct and incorrect predictions. Moreover, while text explanations often come from human experts (e.g., \cite{Jacobs2021, whole2021bansal}), this study enhances the human-centered explainable AI (HCXAI) framework proposed by \citet{humanexp2023morrison} by gathering 156 human-generated explanations (both saliency- and text-based) prior to comparing them with XAI explanations in a comprehensive human-participant study. This methodology enables a deeper exploration of human explanation strategies within the context of a question-answering task, utilizing the SQuAD 1.1v dataset \cite{rajpurkar-etal-2016-squad}.


Our findings show that both performance and subjective experiences with AI are significantly influenced by the accuracy of AI-generated answers. Specifically, explanations for correct answers were perceived as more helpful, of higher quality, and required less mental effort compared to those for incorrect answers. In terms of saliency-based explanations, human-generated explanations were deemed more helpful than those produced by machines (conservative LRP and integrated gradients (IG)). However, when considering general satisfaction with the explanations or trust in the AI across various conditions (encompassing both correct and incorrect answers), these differences between conditions disappeared. This tendency highlights the need to evaluate interaction effects: participants found human saliency maps and answer-only maps to be of higher quality than maps created with the help of IG \textit{for correct predictions}.

While we did not reveal significant differences in performance scores across conditions, we found that there was a strong correlation between trust, satisfaction, and performance scores, indicating that participants who were less trusting of the AI and less satisfied with the explanations generally performed better. While this connection might be supported by  the AI's designed accuracy rate of 50\% (and hence required a certain distrust in the AI to perform well), it might explain why explanations generated with integrated gradients---which achieved the lowest scores for helpfulness, quality, satisfaction, and trust---lead participants to achieve the highest performance of all conditions. Indeed, we found that participants in two control conditions without any explanations did achieve better performance than common explanation types (conservative LRP, human saliency, and ChatGPT explanations). In this context, we discuss the the dilemma of incorrect AI predictions and the dangers of explanation confirmation bias.

We make four contributions in this paper:
\begin{enumerate}
    \item We collected and analyzed 156 human-generated saliency-based and text explanations from an online, crowd-sourcing task ($N=40$). Our analysis show that human saliency-based explanations have little overlap with machine explanations and that human text explanations mostly copied or paraphrased source text (text extractions). 
    \item In a large human-participant study online ($N = 136$), we evaluated different human and machine saliency-based and text explanations. We found that the correctness of AI predictions had a strong, significant effect on all measures (performance, time, quality, helpfulness, and mental effort), machine saliency maps were significantly less helpful than human saliency maps, participants trusted text extractions more than ChatGPT explanations, and that measures of explanation satisfaction, trust in the AI, and explanation helpfulness were negatively correlated with performance scores.
    \item We discuss the dilemma of machine explanations: ``good'' explanations for incorrect AI predictions can lead to negative effects such as over-reliance on AI resulting in decreased performance in collaborative human-AI tasks. Furthermore, we discuss how participants might try to match their intuition of relevance with saliency maps themselves as a heuristic mechanism to evaluate explanations. We discuss the dangers of explanation confirmation bias.
    \item We provide recommendations for designers and researchers to use saliency maps and mitigate the dangers of explanation confirmation bias.
\end{enumerate}


\section{Related Research}
\label{related}

We describe the related work that introduces the need for XAI, research on the difficulties with explaining deep learning systems, and work on evaluating explanations in XAI.


\subsection{Explaining AI}

The field of eXplainable artificial intelligence (XAI), tasked with explaining, or providing insight, into AI models, has grown tremendously since DARPA announced its XAI program in 2017 \cite{gunning2017explainable, xai_tax_2020_arrieta, vilone_notions, robots-anjomshoae-2019}. This growth has introduced a plethora of research papers that have investigated and reviewed the facets and desiderata of ``explainability'' (e.g., interpretability, understandability, or transparency) \cite{miller2018explanation, Sokol_Flach_one_expl_does_not_fit, xai_tax_2020_arrieta, abdul2018trends, vilone_notions, taxonomy-chromik-2020}, distinguished between inherently transparent models (i.e., ``glassbox'' models like decision trees or support vector machines (SVMs)) and uninterpretable ones (i.e., ``opaque'' models like deep neural nets) \cite{lipton_2018, kaur2020interpreting, weld2018intelligibility, tutorials_lai}, and has provided different types of explanation, those that explain the model itself (i.e., global) or its prediction (i.e., local) \cite{doshi-valez_towards_interpret_ml, danilevsky-etal-2020-survey}.  The timing for this development is urgent as decisions made by more powerful AI can be highly consequential for people (e.g., in healthcare diagnoses or financial loan applications) \cite{lipton_2018}. The necessity for explainability is further underscored by legal requirements like the General Data Protection Regulation (GDPR) \cite{gdpr_wachter}, which mandates that AI decisions be interpretable to ensure transparency and accountability.

\subsubsection{Deep Learning}

Through the advent of the field of deep learning \cite{deep_learning}, and its ability to train large neural networks using large amounts of computational resources, much attention has been paid to explaining deep neural nets \cite{ex_deep_2021_samek, samek2017explainable, lipton_2018}. To make these opaque models more explainable, research has provided text explanations, visualizations, and explanations by examples \cite{ex_deep_2021_samek, samek2017explainable, lipton_2018, danilevsky-etal-2020-survey}, or has tried to find simpler surrogate models (e.g., linear models) with similar accuracy/performance but inherent interpretability \cite{ex_deep_2021_samek, danilevsky-etal-2020-survey}. In the realm of natural language processing (NLP), the domain of this study, \citet{danilevsky-etal-2020-survey} list different explanation techniques (e.g., feature importance or surrogate model), operations (e.g., gradient-based saliency or attention), and visualizations (e.g., saliency maps or natural language explanations). 

While attention mechanisms, a key feature of the Transformer architecture \cite{transformer}, have been suggested as a way to inspect more complex models \cite{vig2019multiscale, bahdanau2014neural, ex_deep_2021_samek, pafla2020researching}, the use of attention as explanation remains controversial \cite{wiegreffe-pinter-2019-attention, jain-wallace-2019-attention, vashishth2019attention, understanding-nmt-ding-2017}. However, given the focus on input features and their importance to the overall prediction, research has suggested using saliency methods instead \cite{bastings-filippova-2020-elephant}. These methods rely on gradients (e.g., integrated gradients \cite{integrated_gradients} or layer-wise relevance propagation (LRP) \cite{lrp}) and usually fall into the class of post-hoc explanations that are produced after the model's prediction/generation \cite{ex_deep_2021_samek, danilevsky-etal-2020-survey, ramanishka2017top}. For example, in the domain of NLP, words in a movie review, that contributed positively or negatively to the sentiment of the review, can be highlighted  \cite{arras-etal-2017-explaining}.
In this research, our goal is to evaluate the effectiveness of saliency-based explanations in human-participant studies, besides text explanations, which have become the predominant method to explain large language models \cite{danilevsky-etal-2020-survey}.

\subsection{Difficulties with Explanation in Deep Learning}

Despite these efforts in the field of XAI and the hope that explanation can mitigate human biases \cite{human-biases-expl-wang-2019}, there is doubt that current AI models like neural nets can provide explanations: \citet{lipton_2018} argues that ``today’s predictive models are not capable of reasoning at all'' because AI models like neural networks only model correlation, and thus are unable to provide causal explanations \cite{abdul2018trends, weld2018intelligibility, pearl1, pearl2, lipton_2018}. However, the social sciences, according to \citet{miller2018explanation}, tell us that people desire causes to explain events, not associative or statistical relationships, as is common in the field of XAI. There is some evidence that this mismatch between people's expectation of explanation and XAI explanations can lead to situations where XAI explanations are misused, misinterpreted, and over-trusted especially when facing the danger of information overload \cite{kaur2020interpreting}.

In the context of large language models, the application of traditional XAI techniques is limited due to the models' complex, sub-symbolic nature. Studies such as \citet{manipulating-int-poursabzi-2021} and \citet{Sokol_Flach_one_expl_does_not_fit} focus on simpler, inherently interpretable models (e.g., linear regression or decision trees) that, unlike neural networks, provide transparent explanation mechanisms. Additionally, model-agnostic approaches like SHAP \cite{shap} and LIME \cite{lime}, as well as the weight of evidence framework, encounter difficulties in high-dimensional feature spaces typical of text generation tasks \cite{historical-confalonieri-2021, weight_of_evidence_2021}. \citet{vilone_notions} find that the vast majority of articles using neural nets compared different explanation methodologies for the classification of either text or images, rather than their generation. Against this backdrop, our study evaluates and contrasts human-generated explanations with those produced by machines in a question-answering setting, seeking effective explanation methodologies for text generation.

In addition to some technical concerns around the efficacy of XAI explanations like saliency maps, attention, or LIME/SHAP (e.g., \citet{sanity-checks-adebayo-2018, jain-wallace-2019-attention, fooling-lime-slack-2020}), some research views XAI explanations methods generally more critically \cite{patterns-chromik-2019, pitfalls-ehsan-2021, stop-Rudin-2019, vilone_notions}, especially when moving away from concerns of faithfulness (i.e., is feature importance consistent with the internal decision model?) towards plausibility (i.e., does feature importance align with human intuition?) \cite{ding-koehn-2021-evaluating}. \citet{stop-Rudin-2019} argues that it is not enough to highlight which parts of the input were relevant in producing an output as these ``explanations'' do not show how the inputs were used to compute the output \cite{stop-Rudin-2019}. In a similar vein, \citet{veri_fok_2023} states that explanations based on feature importance may provide some indication of the relevance of an individual feature, but typically do not allow decision makers to verify AI recommendations in decision making tasks.

\citet{stop-Rudin-2019} highlights a critical issue in XAI: explanations often pertain only to correct predictions, potentially misleading users with undue confidence. The necessity of considering AI prediction accuracy in research designs has been emphasized to assess if explanations can aid in discerning AI errors \cite{eval_saliency_user-2020}. Initial findings, such as those by \citet{humanexp2023morrison}, indicate that the perceived helpfulness of human explanations diminishes when associated with incorrect predictions or explanations. Moreover, \citet{whole2021bansal} found that explanations, irrespective of their correctness, could lead to higher acceptance rates of AI predictions, sometimes even enticing users to accept incorrect predictions as shown in studies (e.g., \cite{VANDERWAA2021103404, Jacobs2021, trust_think2021Buccina}). \citet{trustcal2020zhang} argue that local explanations might not be suitable for trust calibration. Our study explores the effectiveness of human versus machine-generated explanations in helping users discern between correct and incorrect AI predictions in text-based applications, like question-answering, by evaluating their impact through subjective and performance metrics relative to AI answer accuracy.

Because of the difficulty in explaining individual predictions, explainability tools have been considered ``a false hope'' to open opaque boxes in AI \cite{false-hope-Ghassemi-2021} and authors have called to stop the use of opaque box models in high-stakes scenarios altogether \cite{stop-Rudin-2019} or to rely on established, randomized trials to determine the effectiveness of AI models \cite{false-hope-Ghassemi-2021}. These calls might be supported by the fact that XAI explanations have been used to deceive people to obstruct or force users to take certain actions \cite{patterns-chromik-2019}, or have caused other, more unintended, downstream effects (like the false sense of confidence mentioned above). \citet{pitfalls-ehsan-2021} name these effects ``explainability pitfalls'' \cite{pitfalls-ehsan-2021}. 

\subsection{Evaluating XAI Explanations \texorpdfstring{and Human Explanations}{and Human Explanations}}

To investigate the efficacy of XAI explanation, there has been a strong call for more human-computer interaction (HCI) studies with participants using realistic and interactive prototypes  \cite{weld2018intelligibility, abdul2018trends, miller2018explanation, yang2020reexamining, kaur2020interpreting, doshi-valez_towards_interpret_ml, vilone_notions, historical-confalonieri-2021, manipulating-int-poursabzi-2021, schuff2023challenges}. Unfortunately, evaluations in the field of AI often rely on proxy scores (e.g., performance measures such as F1) or perturbation analysis (e.g., \citet{samek2017explainable}), rather than human ratings or qualitative methods, to evaluate the quality of explanations \cite{schuff2023challenges}. For example, \citet{schmidt2019quantifying} ``define'' trust to be a combination of accuracy and performance scores and \citet{robot-trust-edmonds-2019} equated the trust participants have in explanations with their ability to predict a robot's next action. 


However, those scores can diverge from human ratings (e.g., perceived consistency) \cite{schuff2023challenges} and fail to capture trust's ``multidisciplinary and multifaceted nature'' \cite{howtoevaluatetrust2021vereshak}. \citet{howtoevaluatetrust2021vereshak} contend that trust, conceptualized as an attitude rooted in vulnerability and positive expectations rather than mere behavior, is inadequately assessed in many studies due to the absence of realistic outcomes and the lack of emphasis on participants' attitudes towards AI systems. They also highlight the potential for priming effects from setting positive expectations and the significant role of first impressions in trust formation \cite{howtoevaluatetrust2021vereshak}, which can, under the guise of transparency, be exploited by untrustworthy AI to deceive participants about its competence and mislead users into accepting flawed outputs \cite{untrustworthy2023Banovic}. To address these challenges and mitigate AI over-reliance, \citet{trust_think2021Buccina} advocate for cognitive forcing functions that prompt users to critically evaluate AI decisions.


Furthermore, XAI explanations are seldom benchmarked against human explanations, including preferred formats like contrastive explanations \cite{miller2018explanation}. Predictions labeled as ``human-generated'' are often viewed as more fair and trustworthy than those labeled as ``AI-generated'' \cite{assessment2023mok}, yet XAI faces challenges in emulating human-like explanations, particularly in generating counterfactuals for text generation (not in classification such as in \citet{counterfactual2020mothilal}) which require complex causal reasoning \cite{pearl1, pearl2}. \citet{humanexp2023morrison} found that causal human explanations can mitigate AI over-reliance and enhance decision-making accuracy, despite the risk of incorrect rationalizations. Unlike previous studies, our research conducts a direct comparison of XAI and human explanations, focusing on saliency maps, and explores the efficacy of human text and ChatGPT explanations.

Methodological issues complicate XAI evaluations, including a frequent omission of explanation design limitations \cite{robots-anjomshoae-2019} and the absence of proper control groups in many studies \cite{miller2018explanation, context-utility-vasilyeva-2017, goals-explanation-mohseni-2021}. A meta-review highlighted this challenge, identifying only nine suitable studies from hundreds due to necessary criteria like measuring human performance without AI or between-subjects designs, and found no significant effect on classification performance when XAI explanations were provided, compared to just AI assistance \cite{meta-xai-schemmer-2022}. The diversity of explanations and metrics adds to the complexity, as effectiveness varies widely \cite{Sokol_Flach_one_expl_does_not_fit, schuff2023challenges}. The integration of human-centered and functionality-grounded evaluation metrics is rare, yet crucial for understanding the discrepancies in XAI effectiveness \cite{eval-expl-zhou-2021, zana-sub-measures-2020}. Our study addresses these complexities by directly comparing human and XAI explanations across multiple question-answering tasks including both correct and incorrect answers, examining the interplay between subjective and performance metrics.

\subsection{\texorpdfstring{Research Motivation}{Research Motivation}}


Our research delves into the disparity between human explanations and state-of-the-art XAI explanations by leveraging human-centred XAI (HCXAI). Following \citet{humanexp2023morrison}'s approach, we first gather human explanations in question-answering tasks to understand what type of explanations humans produce. We then expand on their earlier work and compare these human explanations (human saliency and human text explanations) to different XAI-generated explanations. The aim is to assess and compare the efficacy of human and XAI explanations, taking into consideration the accuracy of the predictions involved.

Furthermore, existing studies, such as \citet{assessment2023mok}'s, show a preference for human-generated predictions over AI's, even when they are identical. However, it's not yet clear how this preference changes when both human and XAI explanations for different predictions are framed as AI-generated.
Moreover, while many studies provide feature-based visualizations or saliency maps (e.g., relevance of certain variables as in \citet{Jacobs2021, trustcal2020zhang}), they are usually focused on classification tasks. To the best of our knowledge, our study describes the first study investigating XAI explanations with incorrect predictions in tasks including text generation like question-answering (\citet{whole2021bansal} use expert explanations for their multiple-choice task and saliency maps for a classification task). Lastly, given the lack of studies that include proper control conditions and both subjective and performative metrics our goal is to thoroughly evaluate current SOTA XAI explanation techniques in text generation.

\section{Study Design}
\label{study_design}

In this study, our goal was to compare the efficacy of human and machine explanations of language predictions. To collect and generate these explanations, we needed to ground our study in a specific task that requires specific explanation design. In this section, we describe our task and its explanation goals, and how we developed appropriate explanations to fulfill those goals. At the end of this section, we list our research questions and provide an overview of the study.

\subsection{Task Design}



\begin{figure*}
\centering
\begin{subfigure}{.5\textwidth}
  \centering
  \includegraphics[width=\textwidth, trim={30cm 81.5cm 32cm 7cm},clip]{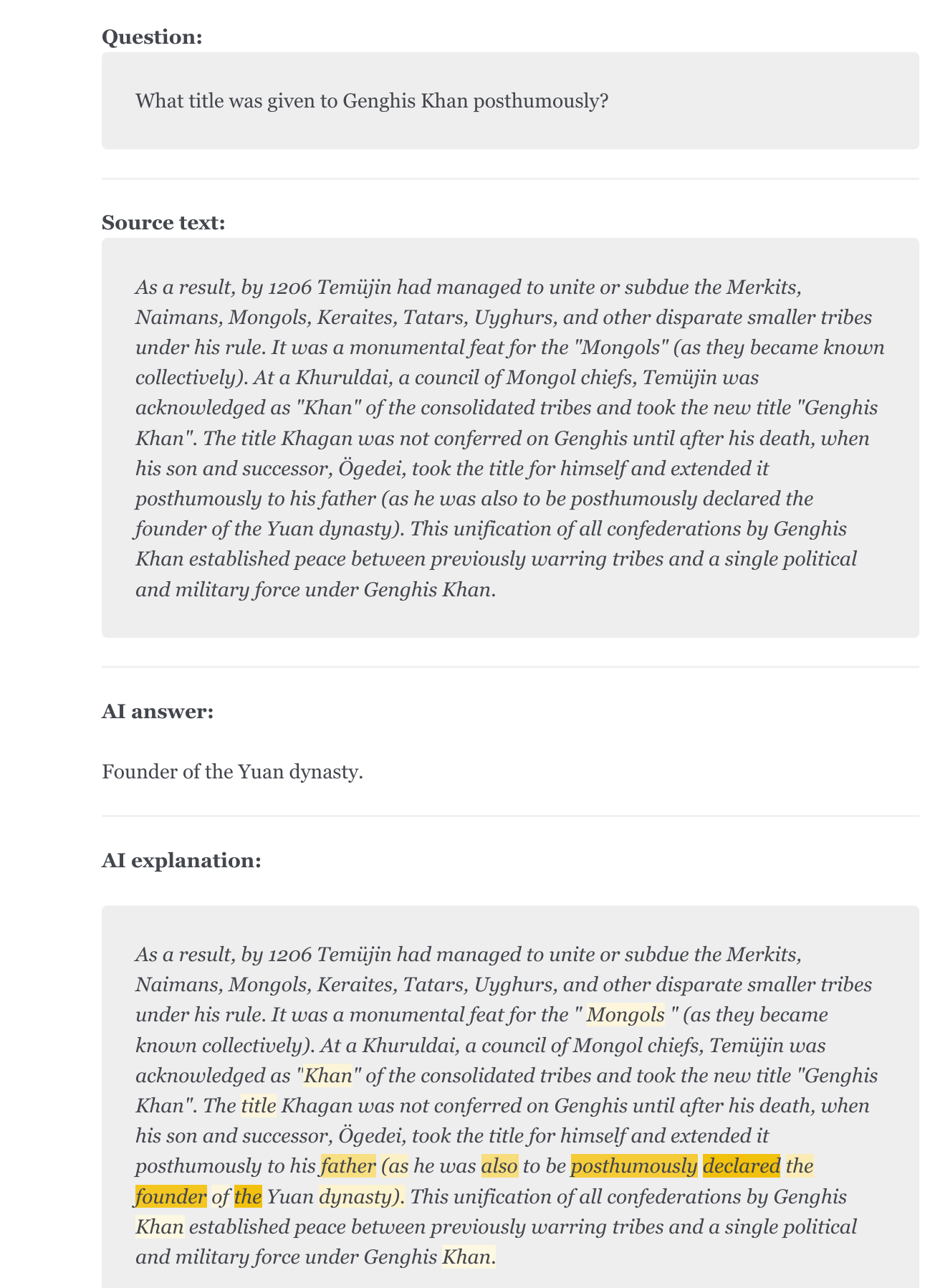}
  \caption{Machine saliency map generated by con-LRP}
  \label{fig:lrp_overview}
\end{subfigure}%
\begin{subfigure}{.5\textwidth}
  \centering
  \includegraphics[width=\textwidth, trim={30cm 81.5cm 32cm 7cm},clip]{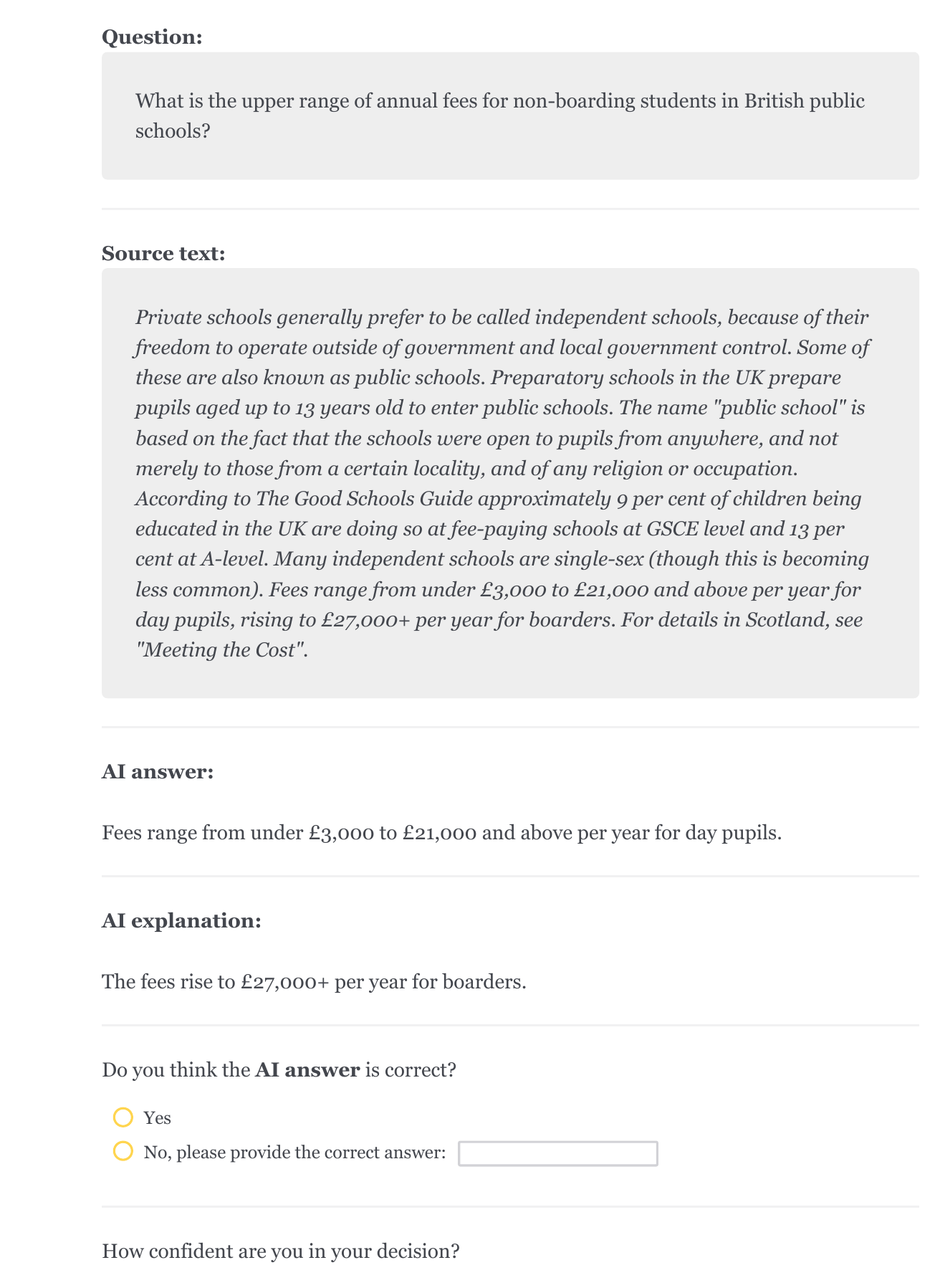}
  \caption{Text explanation provided by human participant}
  \label{fig:expl_overview}
\end{subfigure}
\caption{Screen shots of two question-answering tasks. Participants were exposed to questions, source texts, answers, and explanations, and asked to evaluate the answers with the help of the explanation and source text. Despite presenting both human and machine-generated saliency- and text-based explanations across multiple between-subjects conditions, all participants were informed that the answers were AI-generated. The example on the left featured an answer from Bert, our language model, with a saliency map produced via conservative-LRP, while the right showcased a human-provided answer and contrastive explanation.}
\label{fig:task_overview}
\Description{The figure shows two prints of our question-answering tasks. Each task consists of a question, a source text, an AI answer, and an AI explanation. The task on the left side focuses posthumous title given to Genghis Khan. It includes a source text discussing Temüjin's unification of Mongol tribes and his subsequent acknowledgment as Genghis Khan, explaining that the title "Khagan" was conferred posthumously when his son, Ögedei, extended it to him. The AI's answer to the survey question is "Founder of the Yuan dynasty" which is incorrect. The AI provides an explanation generated by conservative-LRP and highlights parts of the source task that seem to support the answer. Some word highlights across the source text are scattered. The task on the right focuses on the topic of annual fees for students in British public schools. It includes a question about the upper range of these fees for non-boarders, followed by a source text explaining the terminology and statistics related to fee-paying schools in the UK. The answer provides a specific range for day pupil fees. To explain this answer, a human participant contrasted it with the fees of boarders. Thus, the explanation is contrastive. The figure also includes sections asking for the viewer's opinion on the correctness of the AI's answer and their confidence in their decision.
}
\end{figure*}

While explanations in text classification (e.g., sentiment analysis, which predicts text to have either a positive or negative sentiment) have used saliency-based methods to highlight which words contribute most to the classification (e.g., \citet{arras-etal-2017-explaining}), and these methods can help humans understand model predictions by discovering its biases (e.g., \cite{xai-for-transformers}), it becomes more difficult to determine the role and purpose of explanation in text generation. Here, explaining \textit{why} an AI produced a specific text after a prompt entails addressing broader and more open-ended considerations that may extend beyond the immediate task or human-AI interaction, such as the training data's influence on the generated content.

In our work, we focus on question-answering, a popular and challenging benchmark test, where \textit{what} is to be explained (i.e., an answer) is text, rather than a numerical value or a label. Specifically, we rely on the SQuAD 1.1v dataset for study tasks \cite{rajpurkar-etal-2016-squad}, which requires AI and humans to produce an answer given a question about a source text. While the SQuAD is an extractive dataset in which the answer to the question lies directly in the source text, we anticipated the same challenges explanation faces in text generation (e.g., saliency maps not mapping neatly to sentiment scores as in classification). Nonetheless, the extractive nature of answers limits the scope of answers that can be given which can simplify study design. Furthermore, explanations that highlight source text can help participants notice important features which serve as means to verify extracted answers \cite{veri_fok_2023, eval_saliency_user-2020}. An example question and source text (in addition to an AI answer and explanation) of the SQuAD can be found in the task overview in \autoref{fig:lrp_overview}. 



In previous research, explanation goals are often implicit and abstract (e.g., ``explaining model predictions'' or ``explaining model architectures'') or simply focus on performance measures \cite{stop-Rudin-2019, schuff2023challenges}. By grounding our evaluation in question-answering we explicitly formulate our explanation goals (as recommended by \citet{goals-explanation-mohseni-2021}):

\begin{enumerate}
    \item Our first goal is to explain \textit{why the answer to a question is correct}. Consequently, to achieve this verification the focus of our study is on local explanations that can help users determine whether the AI actually produced the right answer and achieve performance which is higher than that of the participant or AI alone \cite{veri_fok_2023}.
    \item Our second goal is \textit{increasing the trust and satisfaction of users}. We expect users are more satisfied with AI explanations and are able to trust AI more if the AI is able to produce good explanations.
\end{enumerate}

\subsection{Explanation Design}

Survey literature proposes various explanation methodologies (e.g., feature-importance, surrogate models, raw-examples, and natural language explanations) for clarifying deep learning models' decisions \cite{guidotti2018black, danilevsky-etal-2020-survey}. However, the suitability of surrogate models and example-driven explanations for large language models and text generation tasks is debatable. Surrogate models, exemplified by LIME, predominantly address classification tasks due to limitations in managing extensive feature spaces \cite{vilone_notions}. Furthermore, large language models are powerful \textit{because of} their size which makes finding a simpler model counter-intuitive \cite{gpt3}. Despite suggestions for using raw examples for neural nets \cite{danilevsky-etal-2020-survey}, we argue these strategies are again more aligned with classification, considering the absence of a structured knowledge base and reliance on unstructured data in our context. Our research incorporates natural language explanations generated via ChatGPT \cite{chatgpt} to study the efficacy of text explanations, yet our primary focus remains on feature-importance explanations like saliency-maps, which are elaborated in the subsequent section.

\subsubsection{Gradient-based explanations}


Considering the debate over the reliability of attention weights \cite{vashishth2019attention, bastings-filippova-2020-elephant} and their potential complexity for non-experts, alongside the risk of information overload in layer-wise attention visualizations \cite{too-much-kuleza-2013}, this study prioritizes gradient-based explanations of feature importance as recommended by \citet{bastings-filippova-2020-elephant}. These are chosen for their simplicity and efficiency relative to methods like surrogate models or occlusion analysis \cite{overview-gradient-based-ancona-2017}.

\citet{overview-gradient-based-ancona-2017} list gradient * input (GI), integrated gradients (IG), layer-wise relevance propagation (LRP), and DeepLIFT as gradient-based explanations in their overview. With this list at hand, the authors developed a unified framework that describes conditions of equivalence between them (e.g., equivalence between LRP and GI \textit{if} Rectified Linear Units (ReLUs) are used). The model we used in our study, BERT \cite{devlin2019bert}, uses Tanh and has additive bias which is why the conditions of equivalence do not hold. Nonetheless, \citet{overview-gradient-based-ancona-2017} find a strong similarity between gradient-based explanations in their evaluations.

We chose to implement conservative-LRP and IG in our study for several reasons. \citet{xai-for-transformers} specifically investigate Transformer models \cite{vashishth2019attention}, the main architecture behind language models like GPT \cite{gpt2, gpt3, chatgpt} and Bert \cite{devlin2019bert}, and find that the layer normalization and the attention heads of the Transformer models break the relevance conservation in LRP which, for example, leaves some attention heads under- or over-represented in the explanation. In their evaluation, the researchers show that their adaptation of LRP, conservative-LRP, achieves better performance compared to GI and other measures across a wide range of benchmarks \cite{xai-for-transformers}. Given that DeepLIFT will likely struggle with the same propagation issues as LRP in Transformers due to the backward pass, we rely on this adaption of LRP by \citet{xai-for-transformers}. Furthermore, we evaluate integrated gradients (IG) as our second saliency-based method as it is a commonly used method that does not rely on full backward pass like LRP or DeepLIFT but it is still able to fulfill useful requirements like the completeness axiom \cite{overview-gradient-based-ancona-2017, integrated_gradients}. Though we are aware that research has found that smoothing algorithms such as smoothgrad \cite{smoothgrad} can increase plausibility in comparison to integrated gradients for transformers \cite{ding-koehn-2021-evaluating}, there is no evidence yet that smoothgrad outperforms IG when considering the entire input prefix (as highlighted by \citet{ding-koehn-2021-evaluating}). Hence, in our study, we do not consider such algorithms.

\subsection{Research Questions}


In this study, our overarching goal is to provide empirical evidence for the efficacy of human and machine text and saliency-based explanations. Thus, our first goal was to collect human explanations and study the kind of explanations humans provide. As a second goal, we wanted to compare collected explanations to SOTA machine explanations (IG, con-LRP, ChatGPT). Finally, once explanations were collected, generated, and analyzed, we were able to compare them empirically in a human-participant study including a wide range of metrics such as trust, curiosity, and satisfaction, in addition to more common performance measures.



With these overall goals, we address the following research questions:

\begin{itemize}
    \item \textbf{RQ1}: What explanations do humans provide when explaining their answers to a question? 
    \item \textbf{RQ2}: How different are human and machine saliency-based explanations?
    \item \textbf{RQ3}: How are human and machine text and saliency-based explanations empirically evaluated in terms of a wide range of factors such as trust, satisfaction, performance, and confidence, when evaluated in a human-participant task including both correct and incorrect answers? 
\end{itemize}
\vspace{5pt}

\begin{figure*}
    \centering
    \includegraphics[width=\textwidth]{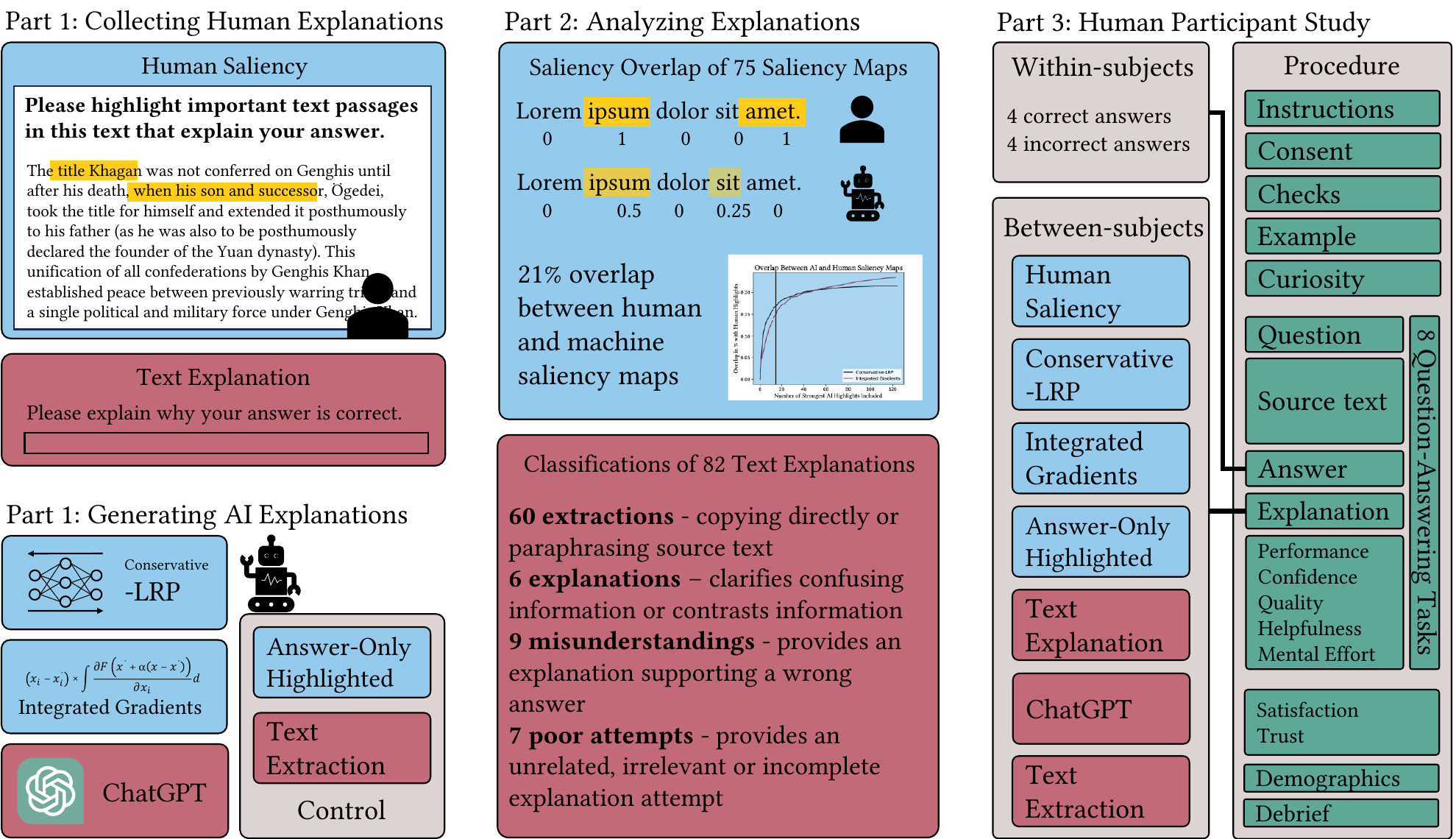}
    \caption{Our study, divided into three parts, aimed to assess the efficacy of various human and machine-generated saliency- and text-based explanations through empirical research. Initially (left section), we gathered human explanations from 40 participants in a crowdsourcing task, where they were tasked with creating saliency-based explanations by highlighting source text and providing text explanations. Subsequently, we utilized techniques such as conservative-LRP, integrated gradients, and ChatGPT to generate machine explanations. Analysis (middle section) revealed a limited overlap (21\%) between human and machine-generated saliency maps, with the latter being more dispersed. Through thematic analysis, we developed a coding scheme to classify human explanations into four categories, which was then independently applied. In the final part (right section), 136 participants evaluated the collected and generated explanations across seven conditions, including four saliency-based and three text-based, including control conditions. This evaluation measured both objective (e.g., performance) and subjective (e.g., satisfaction) metrics, exposing participants to both correct and incorrect answers within a between-subjects design.}
    \Description{The figure displays how our study is divided into three parts. First, when collecting human explanations we ask participants to highlight source text. An example of a source text with highlights is given. We then ask participants for text explanations. An example of a text box is given. At the bottom of this section, we list all the AI and control explanations we generated: conservative-LRP, integrated gradients, ChatGPT explanations. Second, in the middle part of the figure, we give an example of the difference between human and AI attribution scores and provide a figure for the overlap between different AI and human saliency maps. We then list the classifications of text explanations: extractions, clarifications, misunderstandings, and poor attempts. Third, in the right section of the figure, we outlines a complex procedure for the human participant study, including within-subjects and between-subjects procedures, checks, curiosity questions, and measures of performance, confidence, quality, helpfulness, mental effort, satisfaction, trust, along with demographic data collection and a debriefing session.}
    \label{fig:study_diagram}
\end{figure*}

The study consisted of three parts. In the first part, text- and saliency-based explanations were collected from humans and generated using XAI techniques. In the second part, aimed at answering \textbf{RQ1} and \textbf{RQ2}, human text explanations were classified and human and machine saliency maps were compared. In the third part, a new group of participants was asked to evaluate previously collected and generated explanations, allowing us to investigate \textbf{RQ3}. A detailed overview of the study's structure is illustrated in \hyperref[fig:study_diagram]{Figure~\ref*{fig:study_diagram}}.

\section{Part 1: Collecting/Generating Explanations}
\label{sec:collection}

In this section, we describe how we generated XAI explanations and collected human explanations in a crowdsourcing task for our human-participant study described in \autoref{methodology}.

\subsection{Generating XAI Explanations}

\subsubsection{Language Model}
\label{language_model}

We used the large Bert model \cite{devlin2019bert}, an architecture based on the Transformer \cite{transformer}, with around 335 million parameters for our study. We used a pre-trained model that we downloaded from Hugging Face \cite{finetuned_huggingface}. We did not further fine-tune the downloaded model.

\subsubsection{Saliency-based XAI explanations}

We leveraged integrated gradients \cite{integrated_gradients} and conservative LRP \cite{xai-for-transformers} to produce two saliency-based visualizations per question-answering task. The attributions from each method were produced using the code bases of \texttt{pytorch/ captum} \cite{captum} and \texttt{cdpierse/transformers-interpret} \cite{Pierse_Transformers_Interpret_2021} for IG, and \texttt{AmeenAli/XAI\_Transformers} \cite{xai-for-transformers} for LRP. For both methods, we produced the attributions of relevance with respect to the embedding layer (similar to \citet{viz-bert-tut, Pierse_Transformers_Interpret_2021}). The attributions were retrieved for both the start and the end token of the answer (which is the output of the language model). To increase interpretability and make explanations accessible to participants we did not show participants two separate visualizations but averaged the attributions for both tokens. 

We chose to produce attributions with regards to the actual predicted start and end tokens. This decision stands in contrast with \citet{xai-for-transformers} who produce explanations for ground truth labels, which we think biases the evaluation of explanations, as explanations are only shown for correct ground truth answers (which are often not available in deployed AI applications). Furthermore, we use the same composition scheme as \citet{ding-koehn-2021-evaluating} to combine attribution scores of multiple tokens into one when one word is split up into multiple tokens. As the baseline for integrated gradients, we chose 0-vectors for the different embedding layers, similar to \citet{Pierse_Transformers_Interpret_2021}. Furthermore, we averaged the attributions of punctuation tokens (e.g., `.') with the closest letter word. 

\subsubsection{ChatGPT explanations}
\label{chatgpt_expl}

We leveraged ChatGPT to explain the answers of our language model. For each explanation, we provided the source text, the question, and the answer. While we acknowledge that ChatGPT does not generate explanations based on our underlying language model (see \autoref{language_model}), but instead generates explanations on a conceptual text level, we stress that human explanation is often text-based. Our motivation to leverage ChatGPT then was to use the current SOTA language model to produce the best type of machine text-based explanations that we can compare against human text explanations. Generally, our goal was to analyze human explanation strategies, and compare and evaluate different types of machine and human explanations (e.g., saliency-based or text-based). In the supplementary material, we discuss how we derived our prompt to produce ChatGPT explanations. The specific prompt to ChatGPT was:

\begin{quote}
        "You are a question-answering assistant that explains answers to questions. You will be provided with source texts that contain approximately 150 words, a question about these texts, and an answer. Your job is to explain why the provided answer is correct. Provide explanations in one or two sentences. Write the explanations in an instructional tone."
\end{quote}

\subsection{Collecting Human Explanations from Human Participants}

\subsubsection{Participants and remuneration}



The study was executed online, engaging participants from Prolific for a survey administered via Qualtrics. Data from 40 participants were collected in this first part of the study, with ages ranging from 20 to 53 years (median age 27) and all fluent in English. Gender distribution was even, with 20 male and 20 female participants, confirmed by self-reports. Attention and comprehension checks were conducted, leading to the exclusion of eight additional participants for failing these checks, whose data were subsequently discarded. Upon completing the experiment, participants were redirected to Prolific for compensation at a rate of \pounds 9.00 per hour, with the average participation time being 15 minutes.


\subsubsection{Dataset}

For our study, we selected 40 random question-answering (QA) tasks from the SQuAD 1.1v dataset \cite{rajpurkar-etal-2016-squad}, ensuring a mix of correct and incorrect AI responses, with half of the tasks deliberately chosen for their incorrect answers by the AI model. This selection was achieved by evaluating the entire SQuAD validation set with our language model, identifying 364 tasks where the model's answers had an F1 score of 0. Each task was manually reviewed to confirm suitability for participants and accuracy of AI responses, excluding those with problematic questions, divergent ground truth answers, politically sensitive content, or discrepancies between AI responses and scoring metrics.

The SQuAD 1.1v dataset, recognized for its widespread use and relevance in question-answering research as indicated by download metrics on Hugging Face \cite{Wolf2019HuggingFacesTS}, was chosen for its suitability for an online format, with source texts averaging ~200 words. We further refined our selection to tasks with source texts ranging from 125 to 175 words to maintain participant engagement. The human performance on SQuAD 1.1v is 77\% and 86.8\% for exact match metric and the F1 score, respectively.

\subsubsection{Qualtrics survey}

Excerpts of the Qualtrics survey can be seen in \hyperref[fig:study_diagram]{Figure~\ref*{fig:study_diagram}} (left section). When collecting human saliency maps, we displayed the entire source text again and asked participants to highlight text passages that explain participants' answers. This highlighting was done with the help of the ``highlight'' question type provided by Qualtrics. For the collection of text explanations, we simply asked participants to explain why their answer was correct in a text box.


\subsubsection{Procedure}



Participants were recruited via Prolific and directed to a Qualtrics survey, where they were briefed, asked for consent, and underwent instruction, attention, and comprehension checks before being debriefed at the end of the study. Demographic information was also collected. In this study part, participants engaged with four questions and corresponding source texts, tasked with answering and explaining their responses (i.e., why the answer is correct). For each question, they were randomly assigned to either generate text explanations or highlight relevant portions in the source text to explain their answers.

\section{Part 2: Analyzing Collected/Generated Explanations}

In subsequent sections, we address \textbf{RQ1} and \textbf{RQ2} through two approaches: initially conducting a qualitative analysis of human text explanations to categorize the types provided, followed by a comparative examination of human versus machine-generated saliency-based explanations.


We gathered 74 saliency-based and 82 text explanations, totaling 156. Compared to the SQuAD 1.1v official human benchmarks \cite{rajpurkar-etal-2016-squad}, participant performance was lower, with a 59.0\% exact match score and 68.8\% F1 score, versus official scores of 77.0\% and 86.8\%, respectively. This discrepancy may be attributed to our study not encouraging one-word answers, which adversely affects performance on both metrics. Manual validation revealed 88.5\% of participant answers were correct, aligning with official metrics. This corrected accuracy rate will be referenced in subsequent analyses.

The length of explanations was similar for saliency-based and text-based explanations. The average number of words highlighted by participants was 14.57 ($SD = 11.61$). The average number of words in text-based explanations was 15.59 ($SD = 9.62$). We ran two logistic regression models finding no statistical significance between the length of explanation and answer correctness for both saliency-based explanations ($OR=0.9654$, $CI=[0.9152, 1.0184]$, $z = -1.29$, $p = .20$) and text-based explanations ($OR=1.0127$, $CI=[0.9370, 1.0946]$, $z = 0.32$, $p = .75$). Another logistic regression model revealed that there was a small significant effect ($R^2 = 0.0538$, $z = -2.493$, $p = .0127$) between the amount of time participants spent on a question-answering task and the answer correctness ($OR=0.9937$, $CI=[0.9888, 0.9987]$): participants that spent less time on a question had a higher chance of solving it correctly.

\subsection{Qualitative Evaluation of \texorpdfstring{Human Explanation Strategies}{human explanation strategies}}
\label{qual_eval_text}


Adopting \citeauthor{humanexp2023morrison}'s \cite{humanexp2023morrison} human-centered XAI methodology, we conducted a coding reliability thematic analysis (TA) for evaluating human explanation strategies \cite{ref2019braunclarke, Braun2019}, applying two iterations of a coding book to our dataset, diverging from \citet{humanexp2023morrison}'s single iteration. Initially, one researcher analyzed text explanations to identify explanation strategies, leading to a collaborative development of a nine-code scheme. Upon applying these codes, we identified issues of confusion and overlap, prompting a refinement to four mutually exclusive codes for clearer thematic categorization. These four codes were:


\definecolor{shadecolor}{HTML}{ce8892}
\begin{shaded}
    \begin{itemize}
    \item EXTRACTION -- copying source text directly or paraphrasing it; often accompanied by pointing at a text passage (e.g.,  ``in the third sentence it says [...]'')
    \item EXPLANATION -- clarifies confusing or additional information and relates it to the answer or contrasts the correct answer with incorrect ones 
    \item POOR -- provides an unrelated, irrelevant or incomplete explanation attempt 
    \item MISUNDERSTANDING –- provides an explanation supporting an incorrect answer 
\end{itemize}
\end{shaded}
Two raters coded the entire dataset of text explanations ($n = 82$). The level of agreement was 86.6\% (Cohen's $\kappa = .711$, $z = 9.77$, $p < .01$). In total, there was disagreement for eleven instances after coding, which the two raters openly discussed until an agreement was found following the coding reliability approach.


The analysis revealed 60 extractions, 6 explanations, 9 misunderstandings, and 7 poor attempts among the responses, with extractions being the predominant form. This prevalence might reflect participants' preference for efficiency or the limitations of the SQuAD 1.1v dataset in facilitating comprehensive explanations. Indeed, the best type of ``explanation'' might be to point at the answer and state its location in the text (e.g., ``the answer is correct because it says it here''). 


This challenge of generating explanations underscores their contextual nature, particularly in the use of contrastive explanations where participants juxtapose potential but incorrect answers against the correct ones, necessitating imagination to identify plausible alternatives. For example in \autoref{fig:expl_overview}, one participant contrasted the right answer ``£21.000'' with the incorrect, but potential and likely answer ``£27.000'', given context, by explaining that the second value relates to ``borders" \textit{in contrast to} ``non-boarders''.


However, the source texts in the SQuAD 1.1v dataset might not always provide these contrastive cases, participants might not easily see them, or not find them relevant at all. Our coding scheme application allowed us to view all participant responses, including incorrect ones, facilitating the identification of possible contrastive answers. The effectiveness of these identified contrasts as explanations ultimately relies on the perception of participants in study part 3.


Evaluations of explanations for incorrect answers revealed two main types: those stemming from clear misunderstandings of the text and those of generally poor quality due to carelessness or time constraints. When reading these poor explanations, it became quite clear that a) these explanations were indeed of poor quality and that b) there is reason to be careful and verify given answers thoroughly. Given the structure of the task, we were not able to identify any deceitful explanations (as there was no motivation to do so).


Among the text explanations collected, half did not directly address the context but rather elucidated on how specific text elements should be interpreted. For instance, a participant noted that a word in parentheses typically signifies its translation, aligning with the question's requirement. This fact highlights that explanations extend beyond mere context, encompassing external interpretative frameworks (such as social or syntactical conventions) that may lie outside of the provided data to the AI.

This qualitative analysis left us with information on what types of explanations (e.g., actual text explanations vs. extractions) participants provide and expect, answering \textbf{RQ1}, and six text explanations that we can use for study part 3. 

\subsection{Comparing Human \texorpdfstring{with Machine Saliency-Maps}{with machine saliency maps}}
\label{saliency_overlap}

In this section, we compare saliency maps and determine the overlap between  collected human saliency maps from study part 1 and machine saliency maps generated by conservative-LRP and integrated gradients. One motivation for this analysis was to deduct guidelines to advance machine saliency maps and improve their appeal for study part 3. For example, we found that machine saliency maps can look scattered and simultaneously highlight words across the entire text input paragraph which is why it might useful to determine a maximum number of words per saliency map that are highlighted. 

Another motivation for this analysis was to develop and compare different overlap metrics that other researchers can utilize when comparing human and machine saliency maps. Humans tend to highlight sparsely and overlap metrics need to account for the fact that overlap metrics can be artificially high for long input texts. To overcome this issue, we only included words that humans highlighted in our overlap calculations. We provide further discussion on overlap metric calculations in the supplementary material.

Mathematically speaking, when defining $x_N$ and $y_N$ as two vectors holding the attribution scores of $N$ words for human and AI saliency maps for a question-answering task, respectively, then we can calculate the overlap only for words that humans highlighted: $mean(1 - abs(x_N - y_N))$ for all $N$ where $x \in X_N: x = 1$. This approach yielded 21.55\% overlap for matched answer correctness and 15.86\% for mismatched answer correctness for conservative-LRP, and 23.68\% and 20.82\% for integrated gradients, respectively, answering \textbf{RQ2}.

\begin{figure}
    \centering
    \includegraphics[width=0.47\textwidth]{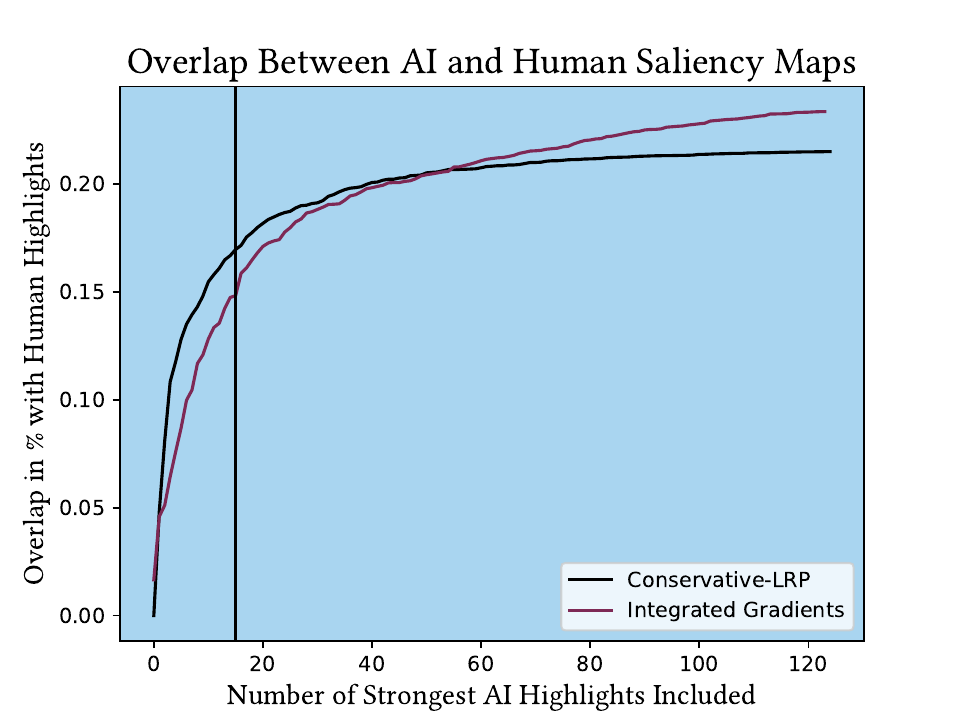}
    \caption{The overlap between AI-generated and human saliency maps varies with the number of AI attributions considered in the analysis. Techniques such as conservative-LRP and integrated gradients assign scores to each word, with full inclusion resulting in the greatest overlap with human maps. However, to minimize explanation clutter and align the visualization of attribution scores with the human average (approximately 15 words, indicated by the blue line), it is observed that the marginal benefit of including additional attributions beyond the first 15 diminishes, leading to a plateau in overlap. Consequently, for empirical evaluation of machine-generated saliency maps, only the 15 most significant attribution scores were visualized.}
    \label{fig:overlap_by_n_highlighted}
    \Description{We plot the relationship between the number of strongest AI highlights included in the saliency overlap calculations with regard to the overlap to human saliency maps. We find that for both con-LRP and IG there is reverse log relationship where overlap does not reach more than 21 percent and plateaus after 15 words. The curve for conservative-LRP rises faster but plateaus earlier. The overlap for integrated gradients is higher if including all words in the overlap calculations.}
\end{figure}

However, these results only hold when including all AI attribution scores for all words in the overlap calculation. Analysis indicated, as shown in \autoref{fig:overlap_by_n_highlighted}, that conservative-LRP demonstrates greater overlap with human maps than integrated gradients when accounting for a limited number of top attribution scores (e.g., for 15 words). Given the average of 14.57 words highlighted by humans in this study and the diminishing returns in overlap beyond 15 words for conservative-LRP, we opted to present only the top 15 attribution scores for both conservative-LRP and integrated gradients in subsequent parts of the study. With this decision, our hope was to make AI saliency maps more ``human-like''.

\section{Part 3: Human Participant Study}
\label{methodology}

After having collected, generated, and analyzed explanations, our goal was to evaluate human- and machine-generated explanations empirically in a human-participant study using a mixed design, addressing \textbf{RQ3} in this study part. We describe the methodology of our human participant study in this section. The study site, remuneration, and underlying dataset (i.e., the SQuAD 1.1v dataset) are the same as in \hyperref[sec:collection]{study part 1 (collection of human explanations)}.

\subsection{Participants}


In this study, data from 136 participants, aged 18-61 (median age 28) and fluent in English, were analyzed. The gender distribution was balanced via Prolific, with 66 female, 67 male, and 3 non-binary participants. Attention and comprehension checks were conducted, excluding 23 individuals who failed the attention check and three who failed the comprehension check, with their data being discarded. Participants were compensated through Prolific at a rate of \pounds 6.68 per hour, with a median participation duration of 22 minutes. Those who correctly answered all eight questions received a bonus of \pounds 0.50, aiming to create a sense of ``vulnerability'' and emphasize the importance of their decisions (see \citet{howtoevaluatetrust2021vereshak}).

\subsection{Study Design}


Our study employed a 9 \textsc{Explanation Type} (between-subjects) $\times$ 2 \textsc{Answer Correctness} (within-subjects) mixed design, ensuring a minimum of 15 participants for each between-subjects condition. \textsc{Answer Correctness} determined the correctness of answers provided by our language model (for machine and control explanations) or human participants (for human explanations), with explanations always aligning with the answer's correctness to avoid mismatched explanations (i.e., we did not provide incorrect explanations to correct answers and vice-versa). Each participant encountered an equal distribution of four correct and four incorrect answers, without being informed of the AI's 50\% accuracy rate to focus on explanation evaluation without priming, adhering to guidelines by \citet{howtoevaluatetrust2021vereshak}.

\textsc{Explanation Type} captures the explanation type participants were exposed to, three of which were text and four of which were saliency-based explanations (and one control in each group for a total of nine for performance and time measures only):
\definecolor{shadecolor}{HTML}{a9d5f0}
\begin{shaded}
Saliency-based explanations:
    \begin{itemize}
        \item \textbf{Conservative LRP/Con-LRP (Machine)} \cite{xai-for-transformers} describes the SOTA method to produce saliency maps for Transformers.
        \item \textbf{Integrated Gradients/IG (Machine)} \cite{integrated_gradients} are a common method to produce saliency maps for neural nets.
        \item \textbf{Human saliency}-based explanations are the highlights participants provided in study part 1.
        \item \textbf{Answer-only (Control)} explanations simply highlight the AI answer in the text. This condition serves as a control condition for other types of saliency-based conditions.
        \item \textbf{No saliency (Control)} is provided to participants to determine the correctness of AI answers. This condition serves as a way to compare performance scores and time commitment.
    \end{itemize}
\end{shaded}
\definecolor{shadecolor}{HTML}{ce8892}
\begin{shaded}
Text explanations:
    \begin{itemize}
        \item \textbf{Text explanations} were provided by previous participants in study part 1. These explanations clarify confusing or extra information in the source text or contrast the correct answer with other potential answers that are incorrect. For text explanations that supported an incorrect answer, we selected explanations tagged as misunderstandings (see \autoref{qual_eval_text}; we did not select poor explanation attempts). 
        \item \textbf{ChatGPT (Machine)} are explanations provided by ChatGPT \cite{chatgpt}.  The exact prompt we used to produce these explanations can be found in \autoref{chatgpt_expl}.
        \item \textbf{Text extractions (Control)} are sentences in the source text supporting AI answers. These extractions are usually text copied from the source text directly, or paraphrased, and usually point to specific passages in the text, similar to how human participants provided text extractions in study part 1. This condition serves as a control condition for other types of text-based conditions. 
        \item \textbf{No explanation (Control)} is provided to participants to determine the correctness of AI answers. This condition serves as a way to compare performance scores and time commitment.
    \end{itemize}
\end{shaded}

\subsection{Selecting Question-Answering Tasks and Explanations for Human Participant Study}

In this study, we tailored our selection of question-answering tasks to include both human and machine explanations, aiming for a full factorial design that encompasses correct and incorrect answers for each explanation type. This necessitated a selective approach due to a limited number of text explanations and incorrect answers provided by participants, coupled with the use of a single AI model producing a singular outcome per question. To achieve a balanced design, we segregated two sets of four questions each for saliency-based and text-based explanations, mindful that incorrect answers might differ across conditions due to variations in human and AI answers. For text-based tasks, selection was straightforward as only four tasks included both text explanations and misunderstandings. On the other hand, for saliency-based explanations related to incorrect answers, we matched four human-generated saliency maps with AI saliency maps, and for correct answers, we randomly chose four from twenty tasks humans and the AI model had answered correctly, ensuring a balanced dataset that allows for an in-depth analysis of explanation types across across correct and incorrect answer instances.





\subsection{Dependent Variables}

\subsubsection{Objective performance measures}

We define two dependent variables to measure performance: \textsc{Performance} and \textsc{Time}. \textsc{Performance} is a binary variable that captures whether participants correctly assessed whether the answer the AI provided is correct or not. Participants were not informed about AI correctness (and they could not assess the correctness of their own assessment). \textsc{Time} is a numerical variable that captures the amount of time in seconds participants spent on each question-answering task. 

\subsubsection{Subjective affect scales and measures}

Following introductory instructions and an explanation example, participants' curiosity was gauged using a modified, unvalidated checklist from \citet{hoffman_metrics} incorporating a 5-point Likert scale. For each question-answering task, participants evaluated their confidence they have in their AI answer assessment on a 0-100 scale, the explanation's quality and helpfulness on 5-point Likert scales following \citet{eval-int-lage-2019} and \citet{schuff2023challenges} respectively, and mental effort using the 9-point Likert validated PAAS scale \cite{paas}. Post-evaluation of eight AI answers, participants rated their satisfaction with the explanations using a validated 8-item satisfaction scale and their trust in the AI on an unvalidated 8-item scale on a 5-point Likert scale by \citet{hoffman_metrics}, excluding the 5th item (``The [tool] is efficient in that it works very quickly.'') due to immediate display of AI responses. Average scores for satisfaction, trust, and curiosity were calculated. 

The choice of scales struck a balance between encompassing a broad spectrum of metrics outlined by \textbf{RQ3} and maintaining an appropriate duration for participant involvement in the experiment, with further details provided in our supplementary materials. The scales from \citet{doshi-valez_towards_interpret_ml}, \citet{schuff2023challenges}, and \citet{paas}, covering quality, helpfulness, and mental effort with concise items, were apt for repeated use across various question-answering task rounds. Notably, \citet{schuff2023challenges} offered helpfulness metrics specifically crafted for question-answering contexts. Given our explanations goals to center user experience and our focus on local explanations, we applied the satisfaction scale by \citet{hoffman_metrics}, which posits that satisfaction is a contextual and retrospective assessment made by users regarding explanations. Finally, we opted for the trust scale by \citet{hoffman_metrics} over  the lengthy human-computer trust (HCT) scale \cite{madsen2000measuring}, which, although unvalidated, amalgamates elements from several validated scales, including HCT and \citet{jian2000foundations}'s trust scale, and is tailored for the XAI context.

\subsection{Deception}


All participants were informed that both answers and explanations originated from AI, even if they were provided by humans. The purpose of this deception was to evaluate the effectiveness of various explanation types without biases related to their origin (e.g., human vs. machine). Given that XAI explanations lack contrastive or causal explanations that humans can provide, and that AI-generated predictions are often viewed as less fair and trustworthy than those from humans, employing deception was crucial to mitigate these biases. The impact of this deception was considered minimal, with participants undergoing debriefing post-study and consent being reaffirmed.

\subsection{Procedure}



Study participants followed the same procedure as participants in study part 1: information letter, consent form, study instructions, two attention checks, and a task comprehension check. At the end of the study they were debriefed about the study and informed if they had been deceived. Furthermore, we collected simple demographic information (i.e., age and gender). The AI was introduced in a neutral tone to avoid positively skewed expectations and reduce bias in the evaluations of explanations \cite{howtoevaluatetrust2021vereshak}. Before encountering question-answering tasks, we asked participants to evaluate their curiosity in the AI and explanations. 


In this third study part, participants encountered eight questions alongside corresponding source texts, answers, and explanations dependent on \textsc{Explanation Type} in a between-subjects design, with the pretense that all responses and explanations were AI-generated. Their primary task was to assess the correctness of each answer with the aid of the provided explanation. Following each question, participants rated their confidence in their assessment of the AI answer, the explanation's quality, its helpfulness in assessing answer correctness, and the mental effort required for evaluation. Finally, we ask participants to fill out the explanation satisfaction and trust scale. A detailed procedure is depicted in \autoref{fig:study_diagram}.

\subsection{Hypotheses}


To the best of our knowledge, this was the first study to compare human saliency maps to AI saliency maps, with no prior knowledge on their relative impact on performance, trust, and satisfaction. We hypothesized that participants would favor human over AI explanations, regardless of their format (text or saliency-based), and posited that any form of explanation would outperform control conditions (answer-only and text extractions), despite some contradictory evidence (e.g., \cite{kaur2020interpreting, meta-xai-schemmer-2022}). "Better" outcomes were defined across dimensions of satisfaction and trust, performance and time, and quality, helpfulness, and mental effort. Our hypotheses were organized around these metrics, instrumentalizing \textbf{RQ3}.

\begin{itemize}
    \item \textbf{Human saliency, IG, Con-LRP > Control}: Participants in human saliency, integrated gradients and conservative-LRP conditions reach higher levels of satisfaction, trust, performance, quality, and helpfulness, and require less time \& mental effort to solve the task than in the saliency-based control condition. 
    \item \textbf{Human saliency > IG, Con-LRP}: Participants in the human saliency condition reach higher levels of satisfaction, trust, performance, quality, and helpfulness, and require less time \& mental effort to solve the task than in the integrated gradients and conservative-LRP condition. 
    \item \textbf{Text explanation, ChatGPT > Text extraction}: Participants in the text explanations and ChatGPT explanations reach higher levels of satisfaction, trust, performance, quality, and helpfulness, and require less time \& mental effort to solve the task than the text extraction control condition. 
    \item \textbf{Text explanation > ChatGPT}: Participants in the text explanations conditions reach higher levels of satisfaction, trust, performance, quality, and helpfulness, and require less time \& mental effort to solve the task than in the ChatGPT explanations. 
\end{itemize}

\begin{figure*}
    \centering
    \includegraphics[width=\textwidth]{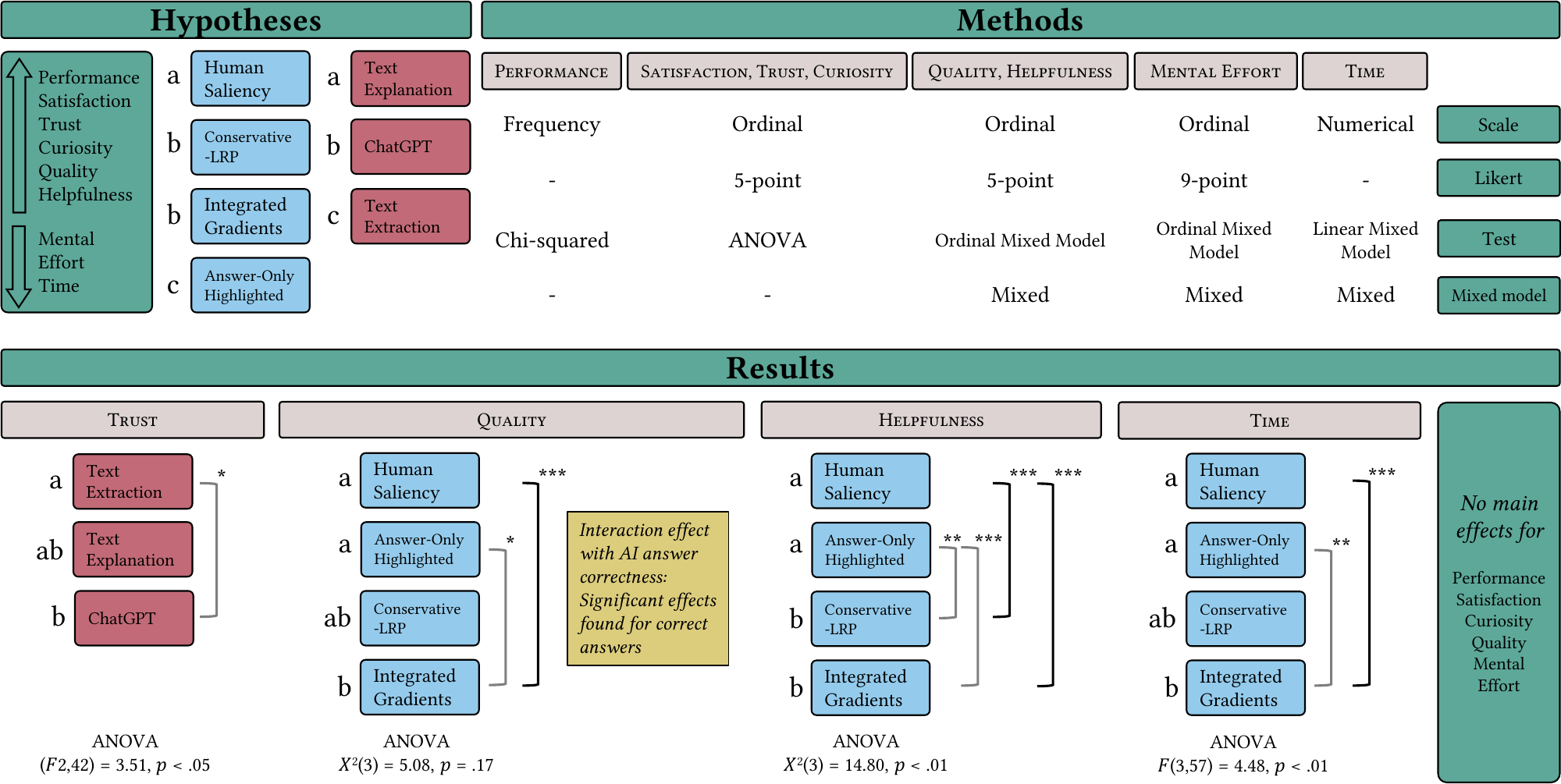}
    \caption{Connection between hypotheses, scales, tests, and results. We hypothesize (top left corner) that, for each of the measures (performance, satisfaction, etc.), the existence of three groups (a, b, and c) that include conditions who are significantly different to all conditions in all other groups on this measure. 
    In the top right, we provide a table that includes basic information about the measures we used in this study including the scale of the measure, whether it is a Likert scale, what test we applied to hypotheses of the measure, and whether we ran a mixed model for the measure which included repeated measurements from participants (i.e., eight question-answering tasks). While we were not able to confirm most of our hypotheses, we represent the most interesting findings of the study: there was a significant difference for trust for text-based explanations, and significant differences for quality, helpfulness, and time for saliency-based explanations. We present significant effects that partially confirm or contradict our hypotheses with the help of black and grey parentheses, respectively.}
    \label{fig:test_connections}
    \Description{We show the connection between hypotheses, scales, tests, and results. We hypothesize (top left corner) that, for each of the measures (performance, satisfaction, trust, curiosity, quality, helpfulness, mental effort, time), the existence of three groups (a, b, and c) that include conditions who are significantly different to all conditions in all other groups on this measure. We display saliency-based conditions as human-saliency in group a, conservative-LRP and IG in group b, and answer-only in group c. For text explanations, we hypothesize that text explanations is in group a, ChatGPT explanations is in group b, and text extractions are in group c. In the top right, we provide a table that includes basic information about the measures we used in this study including the scale of the measure, whether it is a Likert scale, what test we applied to hypotheses of the measure, and whether we ran a mixed model for the measure which included repeated measurements from participants (i.e., eight question-answering tasks). For performance, we list a frequency scale, and chi-squared test. For Satisfaction, Trust, Curiosity, we list an ordinal scale, a 5-point Likert scale, and an ANOVA test. For Quality, Helpfulness, we list an an ordinal scale, a 5-point Likert scale, and an ordinal mixed model. For mental effort, we list an an ordinal scale, a 9-point Likert scale, and an ordinal mixed model. For time, we list a numerical scale, and a linear mixed model. While we were not able to confirm most of our hypotheses, we represent the most interesting findings of the study: there was a significant difference for trust for text-based explanations, and significant differences for quality, helpfulness, and time for saliency-based explanations. For quality for correct answers, we confirmed that human saliency was of higher quality than IG, but answer-only was also of higher quality than IG. For helpfulness, we confirmed that human saliency was of higher quality than conservative-LRP and IG, but answer-only was also more helpful than the machine saliency maps. Finally, for time, we confirmed that human saliency was more efficient than IG, but so was answer-only. We did not find main effects for performance, satisfaction, curiosity, quality, and mental effort.}
\end{figure*}

In \hyperref[fig:test_connections]{Figure~\ref*{fig:test_connections}}, we represent our hypotheses, analysis methods, and results in the same diagram to clarify the connection between them. To describe both our hypothesized and measured differences in the study, we use alphabetic letters to represent groups where group ``a'' is more (less for mental effort and time) than group ``b'' which is in turn more (less for mental effort and time) than group ``c''. Whenever there is a significant difference between two conditions, those two conditions cannot appear in the same group. In our analysis, if there is a third condition with no significant difference to either of the two, mutually-significant conditions being in two different groups (e.g., ``a'' and ``b''), this condition would be in a group with a combined label (e.g., ``ab'').

For example, for saliency-based explanations we define three groups (a, b, and c) and hypothesize that human saliency will be in the first group (group a), con-LRP and IG will be in the second (group b), the the control condition will be in the last group (group c). We hypothesize that there will be a significant difference between all conditions in different groups: human saliency will be significantly different from con-LRP and IG, and human-saliency, con-LRP, and IG will be different from the control condition. This exact logic is represented in the first two hypotheses we define above in text. Similarly, we define our hypotheses with text-based explanations with three groups. We define these groups of hypotheses for each of the scales we measured (performance, satisfaction, etc.).


\subsubsection{Primary results and discussion}

As readers will note in the next section, the results did mostly not confirm our hypotheses. Rather, we found that control conditions performed as well as the best saliency maps (human saliency) and text explanations while there were no significant main effects of \textsc{Explanation Type} for saliency-based explanations on measures of satisfaction and trust. Hence, we explore our data to find out why promising explanation types (e.g., conservative-LRP, human saliency, and ChatGPT) under-performed. 

We found some interesting effects that we illustrate in \hyperref[fig:test_connections]{Figure~\ref*{fig:test_connections}} and further describe in the next section: for text-based explanations, we found a significant difference for trust between conditions, and for saliency-based explanations we found a significant effect for quality of correct answers (interaction), for helpfulness, and time. For these measures, we found that there was no significant difference between human saliency and the control condition, which is why they were grouped together in group ``a'',  while integrated gradients was significantly different from this group and hence was in its own group ``b''.


In section 8, we discuss the nuanced differences in perceived helpfulness and time efficiency between human and machine-generated saliency maps, despite the overall lack of significant effects on trust and satisfaction. These findings emerged from analyses incorporating mixed models that accounted for both correct and incorrect answers. We emphasize the significance of including variables like \textsc{Answer Correctness} in study designs and its influence on both objective and subjective evaluation metrics. Furthermore, we explore the challenge and dilemma of crafting effective explanations for incorrect AI predictions and the tendency for explanation confirmation bias among participants when assessing explanations.

\section{Part 3: Results of Empirical Study}
\label{results}

In this section, we present the results of our human-participant study, targeting \textbf{RQ3}. We compare human and machine text and saliency-based explanation on a wide range of factors, organizing the ensuing subsections accordingly. Because participants were prompted with the same question-answer tasks for all saliency-based explanations, and a different set was used for all text explanations, we perform separate analyses on each of these groups of explanation types. An overview of all the measures taken and what tests were used for each of them can be found in \hyperref[fig:test_connections]{Figure~\ref*{fig:test_connections}}.

\begin{figure*}
    \centering
    \includegraphics[width=.923\textwidth]{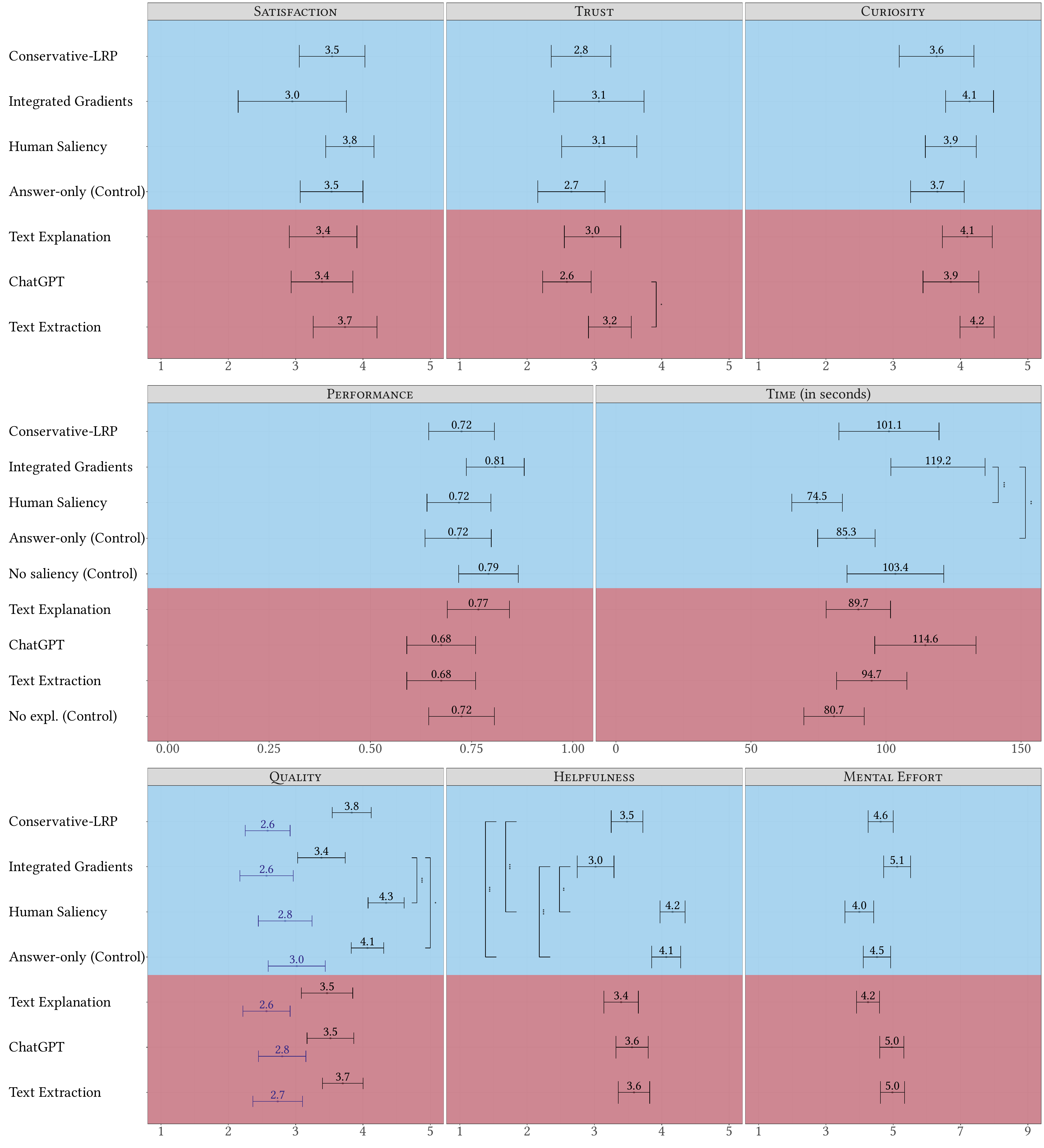}
    \caption{We present the main results of our study for saliency-maps (light blue background) and text-based explanations (red background). For each of our between-subjects conditions, we present the mean and 95\% confidence intervals (CIs) for satisfaction, trust, and curiosity in the first row, performance and time in the second row, and quality, helpfulness and mental effort in the third row. In the second row, we include measures for two extra control conditions, No saliency (Control) and No explanation (Control), in which participants evaluated the same set of AI answers in a question-answering task than in the other conditions, but without any explanation. In the last row, we display the means and CIs for both incorrect (blue) and correct (black) AI answers. As can been in the plots, non-overlapping CIs indicate significant effects: between text extraction and ChatGPT for trust, between human and machine saliency maps for helpfulness, between human saliency maps and integrated gradients for time.}
    \label{fig:plots}
    \Description{For each of our between-subjects conditions, we present the mean and 95\% confidence intervals (CIs) for satisfaction, trust, and curiosity in the first row, performance and time in the second row, and quality, helpfulness and mental effort in the third row. In the second row, we include measures for two extra control conditions, No saliency (Control) and No explanation (Control), in which participants evaluated the same set of AI answers in a question-answering task than in the other conditions, but without any explanation. The control conditions perform as well or better than other conditions. There are overlap between confidence intervals between text extraction and ChatGPT for trust, between human and machine saliency maps for helpfulness, between integrated gradients and human saliency maps for time. Whether the AI was correct or not had a significant effect on time, quality, helpfulness, and mental effort.}
\end{figure*}

\subsection{Satisfaction, Trust, Curiosity}



We performed one-way analyses of variance (ANOVAs) on the measures of satisfaction and trust (measured upon completion of all tasks) and curiosity (measured before all tasks) to compare both the 4 saliency-based explanations and separately to compare the 3 text explanations (\autoref{fig:plots}, top row). All tests showed that these means were not significantly different, except for the trust measure of text-based explanations,
$F_{2,42} = 3.51$, $p < .05$. Pairwise $t$-tests with Bonferroni correction revealed that there was a statistically significant difference between trust averages for text extractions and ChatGPT explanations, $t(27.6) = 2.85$, $p < .05$. 

\begin{table}[tb]
\begin{center}
\begin{tabular}{lcc}
\toprule
          & \multicolumn{2}{c}{\textsc{Performance}}                           \\
AI answer & Correct                     & Incorrect                   \\ \hline
Correct   & \cellcolor[HTML]{C0C0C0}515 & 29                          \\
Incorrect & 283                         & \cellcolor[HTML]{C0C0C0}261 \\
\bottomrule
\end{tabular}
\bigskip
\caption{
Frequency table for performance depending on the correctness of the AI answer (across all conditions). AI correctness significantly influenced outcomes; participants often accepted correct AI answers, while performance dropped to around 50\% for incorrect ones. Grey cells indicate that participants complied with the AI.}
\Description{We provide a frequency table for the performance of participants with regards to whether the AI answer we displayed was correct or not. There is a strong difference in ratios where correct AI answers achieve much higher performance indicating high compliance.}
\label{tab:freq1}
\end{center}
\end{table}  
\begin{table}[tb]
\begin{center}
\begin{tabular}{l!{\vrule width -1pt}c!{\vrule width -1pt}c!{\vrule width -1pt}c}
\toprule
\multicolumn{2}{c}{} & \multicolumn{2}{c}{\textsc{Performance}}\\
{ \textsc{Explanation Type}} & {n} & {Correct} & { Incorrect}\\
\midrule\addlinespace[2pt]
\rowcolor{myblue80}
{Conservative LRP} & 120 & 87 & 33\\[-1pt]
\rowcolor{myblue55}
{Integrated Gradients} & 120 & 97 & 23\\[-1pt]
\rowcolor{myblue80}
{Human saliency} & 128 & 92 & 36\\[-1pt]
\rowcolor{myblue55}
{Answer-only (Control)} & 120 & 86 & 34 \\[-1pt]
\rowcolor{myblue80}
{No explanation (Control)} & 120 & 95 & 25\\[1pt]
\addlinespace[2pt]
\rowcolor{myred80}
{Text explanations} & 120 & 92 & 28\\[-1pt]
\rowcolor{myred55}
{ChatGPT} & 120 & 81 & 39\\[-1pt]
\rowcolor{myred80}
{Text extractions (Control)} & 120 & 81 & 39\\[-1pt]
\rowcolor{myred55}
No explanation (Control) & 120 & 87 & 33\\[1pt]
\bottomrule
\end{tabular}
\bigskip
\caption{
Frequency table for performance split by condition. Despite varying conditions, the ratios of correct to incorrect responses were consistent. Notably, control conditions lacking explanations yielded results comparable to those with the most effective saliency-based and text-based explanations.}
\Description{We provide a frequency table for the performance scores for every condition. Integrated gradients and text explanations performed the best.}
\label{tab:freq2}
\end{center}
\end{table}  



\subsection{Performance (Accuracy)}

\autoref{tab:freq1}, shows the number of times participants evaluated the AI answers correctly (\textsc{Performance}) depending on whether the shown AI answer was correct or not and \autoref{fig:plots} (middle-left) shows these values as a ratio. Whether the AI was correct or not had a highly significant effect on participant performance, $X^2(1)=250.87$, $p<.001$. \autoref{tab:freq2}, shows the number of times participants evaluated the AI answers correctly (\textsc{Performance}) for each explanation condition. As can be seen in the table, the ratios are similar across conditions. Chi-squared tests reveal no significant difference between saliency-based conditions for performance, $X^2(4)=5.08$, $p=.28$ and no significant difference between text-based conditions for performance, $X^2(3)=3.43$, $p=.33$. We note that there is a tendency for control conditions without explanations to perform as well as the best saliency-based (IG) and the best text-based explanations (text explanations). Based on the two control conditions, the question set for text-based explanations (performance = 0.72) is slightly more difficult than the set for saliency-based explanations (performance = 0.79) as can be seen in \autoref{fig:plots} (middle-left).

\subsection{Quality, Helpfulness, Mental Effort, and Time}

Next, we evaluate the effect our explanation conditions and AI answer correctness have on quality, helpfulness, mental effort and time. Since we took these measures after every question (i.e., there are repeated measures for every question for every participant), we ran a mixed-model ANOVA to investigate effects. For quality, helpfulness, and mental effort, we ran an ordinal logistic regression analysis to best model ordinal Likert scales. We present the results in \autoref{fig:within_cond}. The ordinal logistic regression analysis for helpfulness for saliency-based conditions did not satisfy the proportional odds assumption. However, a linear model yielded similar results (albeit the interaction became non-significant). For time, we fitted a linear mixed-model to the log transformed time data. We present the results in \autoref{fig:within_time}. As can be seen in \autoref{fig:within_cond} and \autoref{fig:within_time}, we found that whether the AI answer was correct or not (described by \textsc{Answer Correctness}) impacted participants across all measures for both types of explanations. Participants perceived questions with correct answers to be higher quality, more helpful, lower mental effort, and they took less time.

\begin{table*}
\begin{center}
\begin{tabular}{l!{\vrule width -1pt}l!{\vrule width -1pt}l!{\vrule width -1pt}c!{\vrule width -1pt}r!{\vrule width -1pt}r!{\vrule width -1pt}l}
\toprule
Explanation &{Measure} & {Variable} & {DF} & {$X^2$} & {p} & {Significance}\\
\midrule\addlinespace[2pt]
\rowcolor{myblue80}
Saliency-based & \textsc{Quality} & \textsc{Explanation Type} & 3  & 5.08 & 0.17 &\\[-1pt]
\rowcolor{myblue80}
& & \textsc{Answer Correctness} & 1  & 100.73 & $<.001$ & ***\\[-1pt]
\rowcolor{myblue80}
& & \textsc{Explanation Type}:\textsc{Answer Correctness} & 3  & 8.16 & $<.05$ & *\\[1pt]
\rowcolor{myblue55}
& \textsc{Helpfulness} & \textsc{Explanation Type} & 3  & 14.80 & <0.01 & **\\[-1pt]
\rowcolor{myblue55}
& & \textsc{Answer Correctness} & 1 & 73.62 & $<.001$ & ***\\[-1pt]
\rowcolor{myblue55}
& & \textsc{Explanation Type}:\textsc{Answer Correctness} & 3  & 7.08 & 0.07 & \\[1pt]
\rowcolor{myblue80}
& \textsc{Mental Effort} & \textsc{Explanation Type} & 3  & 5.05 & 0.17 & \\[-1pt]
\rowcolor{myblue80}
& & \textsc{Answer Correctness} & 1  & 38.47 & $<.001$ & ***\\[-1pt]
\rowcolor{myblue80}
& & \textsc{Explanation Type}:\textsc{Answer Correctness} & 3  & 5.28 & 0.15 & \\[1pt]
\addlinespace[2pt]
\rowcolor{myred80}
Text-based & \textsc{Quality} & \textsc{Explanation Type} & 2  & 0.31 & 0.86 &\\[-1pt]
\rowcolor{myred80}
& & \textsc{Answer Correctness} & 1  & 47.55 & $<.001$ & ***\\[-1pt]
\rowcolor{myred80}
& & \textsc{Explanation Type}:\textsc{Answer Correctness} & 2  & 0.80 & 0.67 &\\[1pt]
\rowcolor{myred55}
& \textsc{Helpfulness} & \textsc{Explanation Type} & 2  & 0.15 & 0.93 &\\[-1pt]
\rowcolor{myred55}
& & \textsc{Answer Correctness} & 1 & 44.76 & $<.001$ & ***\\[-1pt]
\rowcolor{myred55}
& & \textsc{Explanation Type}:\textsc{Answer Correctness} & 2  & 1.21 & 0.55 & \\[1pt]
\rowcolor{myred80}
& \textsc{Mental Effort} & \textsc{Explanation Type} & 2  & 3.01 & 0.22 & \\[-1pt]
\rowcolor{myred80}
& & \textsc{Answer Correctness} & 1  & 38.19 & $<.001$ & ***\\[-1pt]
\rowcolor{myred80}
& & \textsc{Explanation Type}:\textsc{Answer Correctness} & 2  & 0.80 & 0.67 & \\[1pt]
\end{tabular}
\bigskip
\caption{Mixed-model, type 2, ANOVAs for saliency-based and text-based explanations based on ordinal log regression models. We found that the within-subjects variable \textsc{Answer Correctness} (i.e., was the AI answer that was provided to participants correct or incorrect) had a strong significant effect on all measures (i.e., quality, helpfulness, mental effort) for both types of explanations. The between-subjects variable \textsc{Explanation Type} had a significant effect on helpfulness for saliency-based conditions. We found a significant interaction effects between \textsc{Answer Correctness} and \textsc{Explanation Type} for saliency-based explanations.}
\Description{We display the results of multiple mixed-model ANOVAs for saliency-based and text-based explanations based on ordinal log regression models. We find main effects for answer correctness for all measures, and explanation type for helpfulness and quality. We also find an interaction effect between answer correctness and explanation type for quality for saliency-based explanations.}
\label{fig:within_cond}
\end{center}
\end{table*} 

\begin{table*}
\begin{center}
\begin{tabular}{l!{\vrule width -1pt}l!{\vrule width -1pt}l!{\vrule width -1pt}c!{\vrule width -1pt}r!{\vrule width -1pt}r!{\vrule width -1pt}r!{\vrule width -1pt}l}
\toprule
Explanation & Measure & {Effect} & {DF} & {Df.res} & {F} & {p} & {Significance}\\
\midrule
\addlinespace[2pt]
\rowcolor{myblue80}
Saliency-based & \textsc{Time}& \textsc{Explanation Type} & 3 & 57 & 4.48 & <0.01 & **\\[-1pt]
\rowcolor{myblue80}
&&  \textsc{Answer Correctness} & 1 & 423 & 35.66 & $<.001$ & ***\\[-1pt]
\rowcolor{myblue80}
&&  \textsc{Explanation Type}:\textsc{Answer Correctness} & 3 & 423 & 0.71 & 0.55 & \\[1pt]
\addlinespace[2pt]
\rowcolor{myred80}
Text-based & \textsc{Time} & \textsc{Explanation Type} & 2 & 42 & 0.70 & 0.50 &\\[-1pt]
\rowcolor{myred80}
&&  \textsc{Answer Correctness} & 1 & 312 & 29.36 & $<.001$ & ***\\[-1pt]
\rowcolor{myred80}
&&  \textsc{Explanation Type}:\textsc{Answer Correctness} & 2 & 312 & 0.04 & 0.96 & \\[1pt]
\end{tabular}
\bigskip
\caption{Mixed-model, type 2, ANOVAs for saliency-based and text-based explanations based on linear mixed models. We found that the within-subjects variable \textsc{Answer Correctness} (i.e., was the AI answer that was provided to participants correct or incorrect) had a strong significant effect on time for both types of explanations. The between-subjects variable \textsc{Explanation Type} had a significant effect on time for saliency-based conditions. We found no significant interaction effects between \textsc{Answer Correctness} and \textsc{Explanation Type} for either type of explanation.}
\Description{We display the results of a linear mixed model for time. We found a significant effect of the explanation type for saliency-based explanations. }
\label{fig:within_time}
\end{center}
\end{table*} 

Furthermore, we found a significant main effect for saliency-based explanations on helpfulness and time. Pairwise post-hoc $t$-tests with Bonferroni corrections revealed that there was a significant difference for helpfulness between human saliency and con-LRP, $t(230)=4.48$, $p<.001$, between human saliency and IG, $t(212)=6.85$, $p<.001$, between answer-only and con-LRP, $t(236)=3.63$, $p<.01$, and between answer-only and IG, $t(226)=5.97$, $p<.001$. In short, machine saliency maps were significantly less helpful than human and control saliency maps. Similarly, we found a significant difference for time between human saliency and IG, $t(183)=-4.48$, $p<.001$ and answer-only and IG, $t(197)=-3.29$, $p<.01$. This finding implies that saliency maps generated by integrated gradients required significantly more time from participants than human and control saliency maps to evaluate the AI answer. 

We also found a significant interaction between \textsc{Explanation Type} and \textsc{Answer Correctness} for quality for saliency-based explanations. When looking at the CIs for quality in \autoref{fig:plots} (bottom), the reader can see that, while CIs are aligned vertically for incorrect AI answers (blue), CIs do not all overlap for correct AI answers (black). Post-hoc, pairwise $t$-tests for correct AI answers (black) for saliency-based conditions revealed a significant difference between quality measures for human saliency and IG, $t(112)=4.33$, $p<.001$, and answer-only and IG, $t(104)=3.20$, $p<.05$.

We also checked for two potential confounding variables: the \textsc{Skill} of participants in answering the questions and the \textsc{Difficulty} of the questions. The \textsc{Skill} of participants was measured by their average \textsc{Performance} across all eight questioning-answering task. The \textsc{Difficulty} of a question was measured by the average \textsc{Performance} of all question-takers for that question in the control conditions without any explanation (where explanation was not able to influence the difficulty of the task). When adding both variables as random variables to our models (see supplementary material) we were able to reproduce our results. However, we additionally found that the interaction effect between \textsc{Explanation Type} and \textsc{Answer Correctness} became significant for helpfulness which does not change our subsequent interpretation of the results. Finally, our study design did not allow participants to check the correctness of AI answers after their assessment, excluding the potential confound of participants' knowledge of correctness on measures discussed in this section.

\subsection{Which Factors Are Related to Performance?}

Finally, we investigate the variables that are strongly related to performance. One of the strongest predictors for performance is whether the provided AI were correct or not, $r(1086)=0.48$, $p<.001$. Nonetheless, we find that participants have some intuition about their performance as confidence values mildly correlate with performance, $r(1086)=0.15$, $p<.001$. Even more interesting are the negative correlations between performance and satisfaction, $r(104)=-0.25$, $p<.01$, performance and trust, $r(104)=-0.38$, $p<.001$, and performance and helpfulness, $r(104)=-0.26$, $p<.01$. 
The moderate negative correlation between performance and trust, with higher performance among participants distrusting the AI, was surprising but also reflecting the context of our study where half of the answers provided were incorrect.

\section{Discussion}

The main findings of our study are as follows (answering \textbf{RQ3}):
\begin{itemize}
    \item Participants trusted text extractions more than ChatGPT explanations.
    \item When provided answers were correct, participants performed better and quicker, perceived the explanation to be higher quality and more helpful, and had lower perceived mental effort to evaluate the answer.
    \item Machine saliency maps (con-LRP and IG) were significantly less helpful than human and control saliency maps.
    \item Saliency maps created with IG required significantly more time from participants to evaluate the answer than human or control saliency maps.
    \item Participants found human saliency maps and answer-only maps to be of higher quality than maps created with IG for correct answers, but not for incorrect answers (interaction).
    \item Measures of explanation satisfaction, trust in the AI, and explanation helpfulness were negatively correlated with performance scores.
\end{itemize}

\subsection{\texorpdfstring{Comparing Machine Explanations with Human Explanations}{Comparing Machine Explanations with Human Explanations}}


Human saliency-based explanations received higher evaluations across various metrics compared to machine-generated explanations, particularly in terms of helpfulness, where they were significantly more favored by participants (mean difference of 1.2 on a 5-point Likert scale between the human saliency and integrated gradients conditions). Furthermore, we found a significant effect of saliency-based explanations on time and on the perceived quality of explanations for correct answers where human saliency-based explanations notably reduced task completion time and were perceived as higher quality for correct answers when compared to machine-generated integrated gradients. However, no significant difference was found between human saliency-based explanations and a control condition highlighting the answer sentence. This suggests that while saliency-based explanations may not be the most effective method for explaining answers in question-answering tasks, human-generated saliency explanations are still preferred over machine alternatives for their helpfulness and efficiency. Future research needs to confirm this finding in further studies including other tasks (e.g., more open-ended question-answering or text summarization).


In the evaluation of text-based explanations, a significant variance in trust scores emerged, particularly between text extractions and ChatGPT explanations, without a corresponding difference in satisfaction or performance metrics. As to the origin of this result we can only speculate, but we found that ChatGPT explanations were more verbose, potentially requiring additional time for comprehension, and leading to lower trust scores due to the complexity of information and perceived opportunity for deception. Alternatively, participants could simply have a preference for succinct explanations, as evidenced by the predominance of text extractions in study part 1, indicating a preference for short text ``pointing'' at the right section in the source text. To close this knowledge gap, future research needs to explicitly investigate the impact of the length of text explanations on trust scores.

\subsection{\texorpdfstring{The Importance of Including Incorrect AI Predictions in Study Design}{The Importance of Including Incorrect AI Predictions in Study Design}}


The correctness of AI answers markedly influenced participants' performance, time investment, perceived explanation quality, helpfulness, and mental effort, overshadowing the impact of the study conditions themselves (see \autoref{fig:within_cond}). Notably, while it is expected that incorrect answers would demand more time and cognitive resources, the discovery that answer correctness more significantly affected the perceived quality and helpfulness of explanations than the explanation type itself suggests a tight coupling between AI prediction and explanation that we further discuss below.

Besides the importance of answer correctness as a factor itself, including it for the sake of revealing other effects can be fruitful. In our study, similar to \citet{trust_think2021Buccina}, we learned that participants that found explanations less helpful and had higher distrust in the AI had a higher chance of performing better. Furthermore, we found that human and control saliency maps were perceived as higher quality than maps generated by IG \textit{but only for correct answers}. We also found main effects for helpfulness and time which did not become present before including \textsc{Answer Correctness} as a variable in our mixed models. Thus, including both correct and incorrect predictions when evaluating AI and XAI explanations in mixed models might be crucial for XAI researchers to gain insight.

\subsection{The Dilemma of Explanation: Incorrect Answers}





Besides the strong effects AI answer correctness had on multiple measures, these effects point at a general dilemma in explanation: explanations are perceived as higher quality and more helpful when they support a correct answer, but explanations for incorrect answers can either make people aware of the incorrectness of an answer (e.g., when the explanation is of poor quality) in the best case, or entice people to accept an incorrect answer in the worst case (e.g., when the explanation looks ``credible enough'').

This dilemma is generally understudied in the field of XAI, especially in within-subjects designs with exposure to multiple rounds of AI predictions and post-exposure evaluations. Future research could investigate the effects of ``bad'' explanations for correct answers versus ``good'' explanations for incorrect ones. In our study, we provided explanations in line with the answer (e.g., they would support the answer even if it is incorrect) and found that higher trust in the AI correlated with poorer participant performance, suggesting a lack of thorough evaluation of AI responses. Additionally, a negative relationship between perceived helpfulness and actual performance was observed, indicating that higher trust in AI or perceived helpfulness of explanations could foster over-reliance on AI, as corroborated by other authors  (e.g., \citet{VANDERWAA2021103404, Jacobs2021, trust_think2021Buccina}). This relationship underscores the risk of participants accepting incorrect answers based on convincing explanations, emphasizing the dangers of the dilemma of AI errors and the need for critical engagement with AI-generated content.


In examining the integrated gradients (IG) condition, the influence of saliency-based explanations on trust-related measures such as time and helpfulness (see \citet{howtoevaluatetrust2021vereshak}) becomes evident, thereby revealing the dilemma. IG explanations, found to be less helpful and efficient compared to human-generated ones, paradoxically led to the highest performance levels in the study. We speculate that participants, upon encountering IG explanations deemed moderately helpful (average Likert score of 3 out of 5) and less efficient, engaged in more thorough scrutiny of AI responses, thereby enhancing their task performance. This increased critical assessment could have deterred quick acceptance of incorrect AI answers. Contrarily, had IG explanations been perceived as more helpful, participants might have more readily accepted incorrect answers, as seen in other explanation conditions. This scenario underscores the complex relationship between explanation helpfulness, participant skepticism, and performance accuracy, illustrating a key dilemma in the effectiveness of XAI explanations.


\subsection{Explanation Confirmation Bias}

Employing saliency maps for explaining generative text predictions presents challenges. At best, saliency maps offer users orientation within text corpora and disclose biases to machine learning practitioners \cite{arras-etal-2017-explaining, xai-for-transformers}. At worst, they fail to elucidate the AI prediction process for humans, as documented by \citet{miller2018explanation, lipton_2018, veri_fok_2023}. We suggest that saliency maps might foster and encourage what we call \textbf{explanation confirmation bias} where evaluators' preconceived expectations about the explanation are matched with the actual explanation when evaluating explanations. For example, participants might find that a saliency map highlights the same words that \textit{they would highlight} leading to a positive perception towards the explanation. \citet{humanexp2023morrison} found some evidence for this bias in their study, noting in their qualitative review of comments that explanations could reassure participants and give them confidence if they supported the assessment of participants, independent of whether that assessment was actually correct or not. In contrast, participants critiqued the AI or were incorrectly influenced by the AI if the AI was not in line with their assessment.


When we generated machine explanations (conservative-LRP and integrated gradients) and tried to evaluate these explanations before our study, we displayed the same bias: when the machine-generated saliency map would highlight the same words we would have highlighted \textit{had we been an AI} we considered the explanation to be of high quality. If not, we were disappointed and frustrated. In our study, we found that participants perceived human saliency maps to be of higher quality than saliency maps produced with the help of integrated gradients \textit{but only for correct answers}. This finding suggests a consensus among participants regarding the expected highlighted words, with higher quality attributed to explanations where the saliency maps matched these expectations. The preference for human explanations in cases of correct answers indicates a shared criterion for important words that machine explanations, particularly those created by integrated gradients, fail to consistently meet.

\section{\texorpdfstring{Recommendations}{Recommendations}}

Based on our three study parts, we make the following recommendations for designers and researchers.

\subsection{\texorpdfstring{Recommendations for Designers}{Recommendations for Designers}}

\begin{itemize}
    \item \textbf{Do not call saliency maps explanations.} Given that machine explanations were less helpful than human machine explanations, which in turn were less performant than control conditions with no explanations, the efficacy of saliency maps as explanation in text generation comes under question. Furthermore, given the dangers of explanation confirmation bias that we describe above we recommend not to speak of explanation when providing saliency maps to users.
    \item \textbf{Provide saliency maps to help users explore data.} A saliency map can help people orient themselves in data. In our study, text of up to 175 words might have been too short to provide benefits, but for longer texts, saliency maps might enable efficient evaluation of AI predictions (as a starting point). \citet{atrey2020exploratory} argue that saliency maps can help designers of AI to explore (rather than explain) agent behavior in tasks concerning deep reinforcement learning (RL). Similarly, human users of AI might use saliency maps to explore (rather than explain) data input to the AI in tasks including text generation.
    \item \textbf{Encourage users to thoroughly evaluate AI predictions.} Independent of whether saliency maps are provided to users, encourage users to thoroughly evaluate AI predictions. To that end, \citet{trust_think2021Buccina} provides an excellent discussion on cognitive forcing functions that can help trigger the analytical processing system of users and decrease overreliance on AI. In low stakes environments, it might be possible to forgo explanation mechanisms and forcing functions altogether and instead provide efficient tools to users to correct faulty AI output. In cases where the AI is expected to outperform human decision-makers, AI decisions can strategically be accepted (see \citet{veri_fok_2023} for a discussion on different forms of reliance).
\end{itemize}

\subsection{\texorpdfstring{Recommendations for Researchers}{Recommendations for Researchers}}

\begin{itemize}
    \item \textbf{Compare machine explanations to human explanations.} To thoroughly evaluate XAI explanations we recommend to collect human explanations of the same type. By restricting participants to the same type of explanation (e.g., saliency maps) when collecting explanations, this type matching between human and machine explanations allows researchers to provide a proper baseline to their investigations. For example, in our study we learned that human saliency maps are more helpful than machine saliency maps, but found no significant performance differences between different types of explanations and control conditions.
    \item \textbf{Evaluate machine explanations for incorrect AI predictions.} We highly recommend that incorrect AI predictions are included when evaluating XAI explanations. Ideally, a factorial design is provided in which explanations follow or do not follow correct or incorrect AI predictions. This design allows researchers to investigate the effects of ``good'' explanations with incorrect predictions and ``bad'' explanations with correct predictions on AI compliance rates and whether people exhibit explanation confirmation bias.
\end{itemize}


\section{Limitations}


As every study, our study comes with limitations. As \citet{meta-xai-schemmer-2022} has shown in their meta-analysis, effect sizes for XAI methods vs. AI assistance  on performance can be very small (e.g., standardized mean differences of $0.07$). Our study was too under-powered to find these differences or exclude them (through equivalence tests). Future research needs to have even larger sample sizes to investigate the efficacy of different types of explanations. 

Second, our research was conducted on naive participants that were not informed or trained on what constitutes an explanation and what does not. Specifically, when we ask participants to rate explanations, we show them only one type of explanation in a between-study design which makes it impossible for participants to compare different explanations with each other. Future research needs to explicitly compare different types of explanations and allow participants to contrast them. Understanding the mental models participants hold of tasks, answers, and explanations will give insight into how participants expect explanation to be.


Third, as is the problem with many studies concerning the evaluation of local explanations, we focused on the evaluation of individual explanations rather than the ``socio-technical'' system \cite{zana-sub-measures-2020} that makes up the human-AI team. Specifically, we did not give means to participants to interact with AI and explanations. As such, future research needs to adapt interactive, theory-driven AI such as \citeauthor{counterfactual2020mothilal}'s \cite{counterfactual2020mothilal} counterfactual explanations to text generation. Ideally, human participants would have the ability to ask the AI model why some other potential answer is not the correct answer, forcing the AI to contrast its answer with contrastive cases. With this ability, we might be able to deal with incorrect AI answers more efficiently.

\section{Conclusion}


In our study, we collected 156 human explanations for a question-answering task and compared them to machine-generated explanations in a human-participant study ($N=136$), covering both correct and incorrect answers. Analysis revealed a 21\% overlap between human and machine (conservative-LRP and integrated gradients) saliency maps and humans primarily using text extractions to justify their answers. These extractions, along with text explanations, misunderstandings, and poor explanation attempts, were categorized using a coding scheme by two independent raters. When evaluating explanations empirically, we found that the correctness of the shown answer had a strong, significant effect on all measures (i.e., performance, time, quality, helpfulness, and mental effort), machine saliency maps (con-LRP and IG) were significantly less helpful than human and control saliency maps, participants trusted text extractions more than ChatGPT explanations, and that measures of explanation satisfaction, trust in the AI, and explanation helpfulness were negatively correlated with performance scores. These findings hint at a dilemma in explanation, where effective explanation can actually decrease performance when supporting incorrect answers or predictions. In this context, we discuss the danger of explanation confirmation bias in the evaluation of explanations.

\begin{acks}
  Thank you to Ali Haider Rizvi, Reza Hadi Mogavi, Ville Mäkelä, Stacey Scott, the University of Waterloo’s Touchlab, UW-MAS, EngHCI group, and the Games Institute. This work was made possible by the Natural Sciences and Engineering Research Council of Canada (NSERC Discovery Grant 2016-04422 and 2018-03962).
\end{acks}
\bibliographystyle{ACM-Reference-Format}
\bibliography{sample-base}











\end{document}